\pdfoutput=1
\documentclass{article}

\usepackage{arxiv}
\usepackage[numbers]{natbib}
\usepackage{tikz}
\usepackage[utf8]{inputenc} 
\usepackage[T1]{fontenc}    
\usepackage{url}            
\usepackage{booktabs}       
\usepackage{amsfonts}       
\usepackage{nicefrac}       
\usepackage{microtype}      
\usepackage{lipsum}
\usepackage{graphicx}
\usepackage{graphicx}
\usepackage{newtxtext}
\usepackage{newtxmath}
\usepackage{array,multirow}
\usepackage{hyperref}  
\graphicspath{ {./images/} }

\newcommand{\full}{\protect\mbox{------}}
\newcommand{\dashed}{\protect\mbox{-\ -\ -\ -}}
\newcommand{\RomanNumeralCaps}[1]
\linenumbers

\providecommand\bnabla{\boldsymbol{\nabla}}    

\title{Biglobal resolvent analysis of separated flow over a NACA0012 airfoil}

\author{
  Laura Victoria Rolandi$^1$, Luke Smith$^1$, Michael Amitay$^2$, Vassilios Theofilis$^3$, Kunihiko Taira$^1$ \\
$^1$Department of Mechanical and Aerospace Engineering, University of California, Los Angeles, CA 90095, USA\\
$^2$Department of Mechanical, Aerospace, and Nuclear Engineering, Rensselaer Polytechnic Institute, Troy, NY 12180, USA\\
$^3$Faculty of Aerospace Engineering, Israel Institute of Technology, Haifa 32000, Israel
}

\begin{document}
\maketitle

\begin{abstract}
The effects of Reynolds number across $Re=1000$, $2500$, $5000$, and $10000$ on separated flow over a two-dimensional NACA0012 airfoil at an angle of attack of $\alpha=14^\circ$ are investigated through the biglobal resolvent analysis. We identify modal structures
and energy amplifications over a range of frequency, spanwise wavenumber, and the discount parameter, providing insights across various timescales. Using temporal discounting, we find that the shear layer dynamics dominates over short time horizons, while the wake dynamics becomes the primary amplification mechanism over long time horizons. Spanwise effects also appear over long time horizon, sustained by low frequencies. At a fixed timescale, we investigate the influence of Reynolds number on response and forcing mode structures, as well as the energy gain over different frequencies.  Across all Reynolds numbers, the response modes shift from wake-dominated structures at low frequencies to shear layer-dominated structures at higher frequencies. The frequency at which the dominant mechanism changes is independent of the Reynolds number. The response mode structures show similarities across different Reynolds numbers, with local streamwise wavelengths only depending on frequency. Comparisons at a different angle of attack ($\alpha=9^\circ$) show that the transition from wake to shear layer dynamics with increasing frequency only occurs if the unsteady flow is three-dimensional. We also study the dominant frequencies associated with wake and shear layer dynamics across the angles of attack and Reynolds numbers, and present the characteristic scaling for each mechanism.

\end{abstract}

\section{Introduction}
Understanding and controlling separated flows is important for improving performance of a range of fluid-based systems. Flow separation over a wing can significantly reduce lift and increase drag, leading to diminished aerodynamic efficiency and higher fuel consumption. This issue impacts not only aerodynamic performance but also affects stability and control of the air vehicle, which can compromise its overall safety and the operational effectiveness. 

A helpful tool for the design of flow separation control strategies is the resolvent analysis \citep{trefethen1993hydrodynamic,jovanovic2005componentwise}, which helps to understand the flow characteristics by linearizing the governing equations around a base flow, modeling the nonlinear terms as an external forcing \citep{mckeon2010critical} and transforming the dynamics into an input-output problem \citep{jovanovic2021bypass}. The use of the time-averaged flow as the base flow, with the assumption of statistical stationarity, allows for the extension of resolvent analysis to turbulent flows \citep{mckeon2010critical, yeh2019resolvent,martini2020resolvent,towne2018spectral}.

With resolvent analysis, complex flow fields are decomposed into coherent modal structures over the frequency. These modal structures are the optimal forcing and response modes and provide a detailed picture of how different parts of the flow are sensitive and respond to specific frequencies. The gain quantifies the energy associated with the forcing and response mode pair, indicating their significance within the flow dynamics. By studying this decomposition, it becomes possible to identify the dominant mechanisms that drive the dynamics of the flow. 

Resolvent analysis thus provides valuable insights for understanding flow unsteadiness and informs flow control strategies. It helps predict effective actuation frequencies and highlights the corresponding forcing and response structures ideal for localized actuation \citep{yeh2019resolvent,ribeiro2024triglobal}.
The strength of resolvent analysis also lies in its ability to capture non-normal effects, which appear when the eigenmodes of the linear operator are non-orthogonal. Non-normality can cause transient disturbance growth, even in flows that are linearly stable. By focusing on the most amplified dynamics driven by non-normal interactions, resolvent analysis exposes mechanisms that might not be detected through conventional stability analysis alone. 

Resolvent analysis has been used for various problems, such as boundary layers \citep{dawson2020prediction,nogueira2020resolvent}, turbulent channel flows \citep{moarref2013model,nakashima2017assessment,zhu2024resolvent}, jets  \citep{schmidt2018spectral,pickering2021resolvent} and airfoil wakes \citep{thomareis2018resolvent,symon2019tale,yeh2019resolvent,yeh2020resolvent}. Among the latter, \cite{yeh2019resolvent} investigated the flow over a two-dimensional NACA0012 airfoil at a chord-based Reynolds number $Re=23000$ and two different angles of attack ($\alpha=6^\circ$ and $9^\circ$), revealing a shear layer dominated mechanism for energy
amplification. Moreover, they used the findings from resolvent analysis to explore the capability of a thermal actuator, which introduces time-periodic heat injection, in suppressing stall and enhancing aerodynamic performance.

A higher Reynolds number ($Re=500000$) flow around a two-dimensional NACA0012 airfoil has also been investigated by \cite{yeh2020resolvent}. In this study, a windowed resolvent analysis was used to localize the forcing and response modes in the laminar separation bubble forming on the suction side of the airfoil. Windowed resolvent analysis has also been applied to identify amplification mechanisms driving the two-dimensional transonic buffet at $Re=2000$ \citep{kojima2020resolvent}.
More recently, resolvent-based control strategies have also been employed for three-dimensional separated flows  \citep{ribeiro2024triglobal}, where the use of the optimal forcing modes has shown a reduction in the size of the separation region and the attenuation of the wing tip vortex.

Originally, resolvent analysis was employed on stable base flows, but it has been later extended to account for unstable base flows through the discounted resolvent analysis \citep{jovanovic2004modeling,jovanovic2005componentwise,rolandi2024invitation}. This approach examines the dynamics over a finite-time horizon, rather than the asymptotic behavior, as the present study also considers.
In this work, we investigate the effects of Reynolds numbers of $Re=1000,\;2500,\;5000$ and $10000$ on separated flow over a NACA0012 airfoil, using resolvent analysis. While previous relevant investigations on separated flows around airfoil using biglobal linear analysis have predominantly focused on lower Reynolds numbers \citep{he2017linear, ribeiro2022wing,tamilselvam2022transient, nastro2023global}, or lower angles of attack when increasing the Reynolds number \citep{gupta2023two}, our work aims to address the transitional regime. Specifically, we analyze the linear amplification mechanisms at transitional Reynolds numbers, for which the flow around the airfoil loses its periodicity and becomes highly unsteady. We provide insights into frequency-dependent responses and bridge knowledge gaps regarding how the Reynolds number influences the linear dynamics in the transitional regime. Our analysis centers on identifying dominant flow structures and characterizing their spatial and temporal behavior, with particular attention to how these features evolve with increasing Reynolds numbers. Finding similarities and physics-based scalings is particularly beneficial in the moderate Reynolds number regime, with the potential to uncover underlying physics also present at higher Reynolds numbers commonly encountered in engineering applications.

\begin{figure} 
\centering
\includegraphics[width=0.8\textwidth]{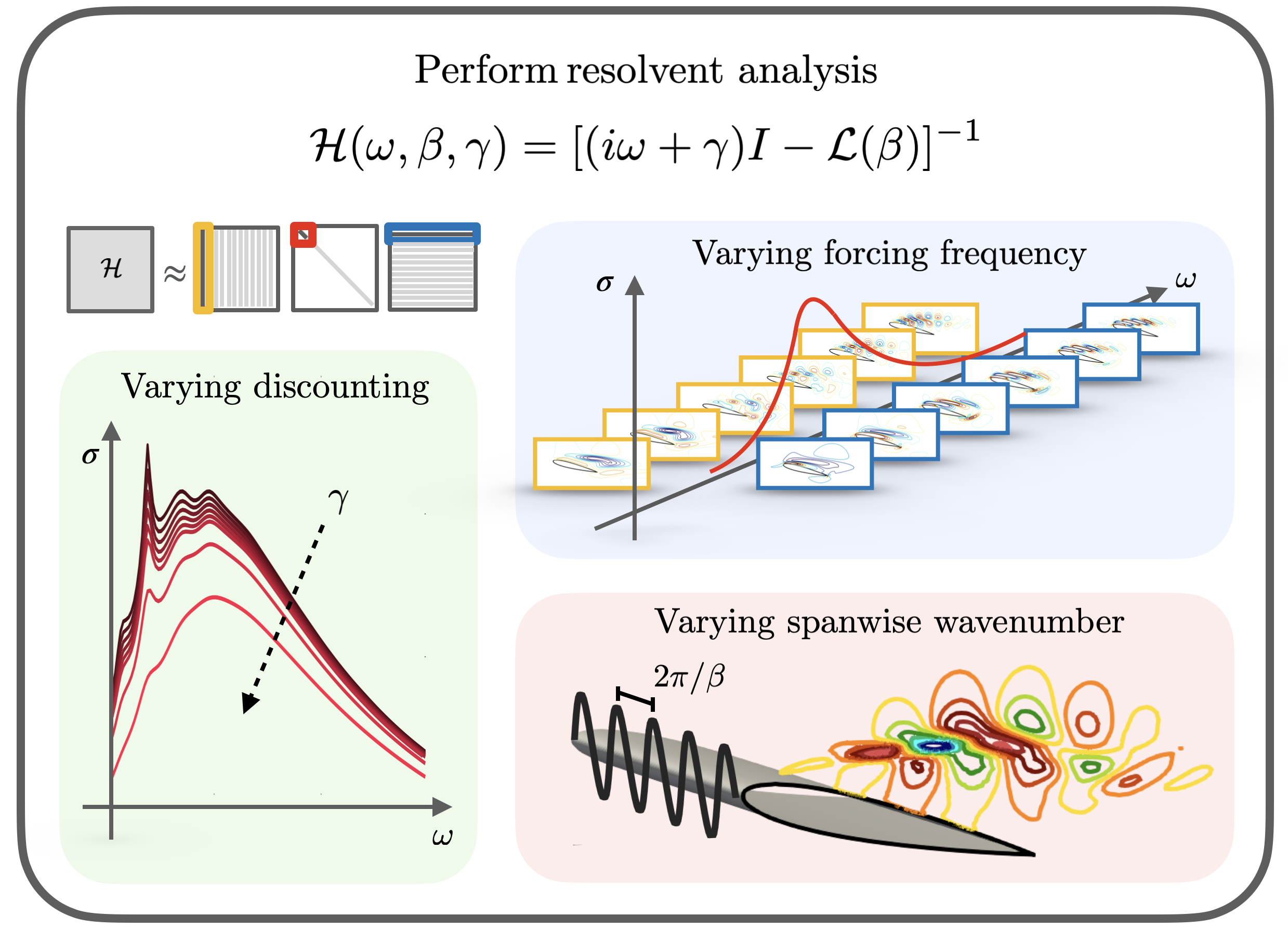}
\caption{Overview of the present resolvent analysis.\label{fig:Overview}}
\end{figure}

The effects of Reynolds numbers are examined in relation to key parameters such as the temporal frequency, the spanwise wavenumber, and the discount parameter, as summarized in Figure~\ref{fig:Overview}, providing a comprehensive understanding of flow dynamics. Investigating the effect of spanwise wavenumber on the dominant flow structures reveals insights into the formation of three-dimensional structures, while analyzing effects of the discount parameter helps determine whether the dynamics evolve over different temporal scales. After introducing the theoretical background and numerical approach in Sect. \S\ref{sec:NumFrem}, Sect. \S\ref{sec:BaseFlow} examines how the Reynolds number affects the separated flow over the airfoil. Sect. \S\ref{sec:ResolventRes} presents the resolvent mode structures and energy amplifications in relation to frequency, spanwise wavenumber, and timescales. Finally, Sect. \S\ref{sec:Normalizations} investigates the impact of a different angle of attack and discusses the scaling behavior of dominant amplification mechanisms with respect to their characteristic frequencies.

\section{Theoretical background and numerical implementation}\label{sec:NumFrem}
We analyze the spanwise periodic flow around a NACA0012 airfoil across various Reynolds numbers. Below, we present the biglobal resolvent analysis theoretical framework and outline the computational setup employed in this study.
\subsection{Biglobal resolvent Analysis}
Let us consider the spatially discretized nonlinear governing equation:
\begin{equation}
\frac{d \mathbf{q}}{d t}=\mathcal{N}(\mathbf{q}),
\label{eq_ns}
\end{equation}
    where $\mathbf{q}=(\rho,\mathbf{u},T)\in \mathbb{R}^{n}$ represents the state vector,  $\rho$ is the density, $\mathbf{u}=(u_x,u_y,u_z)$ is the velocity vector with components along the streamwise ($x$), cross-stream ($y$) and spanwise ($z$) directions and $T$ is the temperature. $\mathcal{N}\in \mathbb{R}^{n\times n}$ is the nonlinear
Navier--Stokes evolution operator,  and $n=N\times 5$, where $N$ is the number of cells in the spatial discretization and $5$ is the number of state variables. We now consider the flow field $\mathbf{q}=\mathbf{q}_b+\mathbf{q}^\prime$ to be composed of the sum of a stationary base flow $\mathbf{q}_b$ and a fluctuating component $\mathbf{q}^\prime$ of small amplitude. The base flow is considered to be the time- and spanwise-averaged flow. By substituting the decomposition in eq.~\ref{eq_ns}, and through a Taylor expansion, we obtain the spatially discretized linearized governing equation for the fluctuating perturbation component, $\mathbf{q}^\prime$:
\begin{equation}
\frac{d \mathbf{q}^\prime}{d t}=\mathcal{L}\mathbf{q}^\prime +\mathbf{f}^\prime.
\label{eq_1}
\end{equation}
Here, $\mathcal{L}\equiv \left. \bnabla_{\mathbf{q}} \mathcal{N} \right|_{\mathbf{q}_b}
    \in\mathbb{R}^{n\times n}$ is the linearized Navier--Stokes operator about the base flow, $\mathbf{q}^\prime=(\rho^\prime,\mathbf{u}^\prime,T^\prime)\in \mathbb{R}^{n}$ is the perturbation and $\mathbf{f}^\prime\in \mathbb{R}^{n}$ collects the nonlinear terms \citep{mckeon2010critical,rolandi2024invitation}.
Due to the temporal and spanwise homogeneity of the base flow, the response ($\mathbf{q}^\prime$) and forcing ($\mathbf{f}^\prime$) can be decomposed through a spatio-temporal Fourier transform as follows
\begin{equation}
    \begin{split}
    \mathbf{q}^\prime(x,y,z,t)=\int^\infty_{-\infty}\int^\infty_{-\infty}\hat{\mathbf{q}}_{\omega,\beta}(x,y)e^{-i\omega t}e^{i\beta z} d\omega d\beta,\\
    \mathbf{f}^\prime(x,y,z,t)=\int^\infty_{-\infty}\int^\infty_{-\infty}\hat{\mathbf{f}}_{\omega,\beta}(x,y)e^{-i\omega t}e^{i\beta z }d\omega d\beta.
\end{split}
\end{equation}
Here, $\beta$ and $\omega$ indicate the spanwise wavenumber and temporal frequency, respectively. By substituting these expressions into eq.~\ref{eq_1}, we find the following input-output relationship,
\begin{equation}
\hat{\mathbf{q}}_{\omega,\beta}(\mathbf{x})=
(-i\omega \mathbf{I}-\mathcal{L}_\beta)^{-1}\hat{\mathbf{f}}_{\omega,\beta}(\mathbf{x})=\mathbf{H}_{\omega,\beta}\hat{\mathbf{f}}_{\omega,\beta}(\mathbf{x}),
\label{eq_2}
\end{equation}
where $\mathbf{H}_{\omega,\beta}\in \mathbb{C}^{n\times n}$ is the resolvent operator that acts as the transfer function between the forcing $\hat{\mathbf{f}}_{\omega,\beta}$ and the response $\hat{\mathbf{q}}_{\omega,\beta}$ at frequency $\omega$ and spanwise wavenumber $\beta$. 
 
 Performing singular value decomposition of the resolvent operator, while retaining only the first $m\ll n$ singular values and right/left singular vectors, we find a low-rank approximation of $\mathbf{H}_{\omega,\beta}$:
\begin{equation}
\mathbf{H}_{\omega,\beta}\approx \mathbf{U}\Sigma \mathbf{V^*}.
\end{equation}
The columns of $\mathbf{U}=[\mathbf{u}_1, \mathbf{u}_2, ..., \mathbf{u}_m]\in \mathbb{C}^{n\times m}$ and $\mathbf{V}=[\mathbf{v}_1, \mathbf{v}_2, ..., \mathbf{v}_m]\in \mathbb{C}^{n\times m}$ hold the response and forcing modes, respectively, while $\Sigma=\text{diag}(\sigma_1,\sigma_2,...,\sigma_m)\in \mathbb{R}^{m}$ retains the gains of the corresponding forcing-response pairs.

The singular value decomposition finds the optimal forcing and response, \textit{i.e.} those that maximize the energy

\begin{equation}
\sigma^2=\max_{\hat{\mathbf{f}}_{\omega,\beta}}\frac{\left<\hat{\mathbf{q}}_{\omega,\beta},\hat{\mathbf{q}}_{\omega,\beta}\right>_{E}}{\left<\hat{\mathbf{f}}_{\omega,\beta},\hat{\mathbf{f}}_{\omega,\beta}\right>_{E}}=\max_{\hat{\mathbf{f}}_{\omega,\beta}}\frac{||\hat{\mathbf{q}}_{\omega,\beta}||^2_{E}}{||\hat{\mathbf{f}}_{\omega,\beta}||^2_{E}},
\end{equation}
where $||\cdot||_{E}$  is a suitable energy norm for the response and the forcing modes. In this work we consider the Chu's norm \citep{chu1965energy,george2011chu}, which is expressed as 
\begin{equation}
    E_{\text{Chu}}=\frac{1}{2}\int_V\left(\bar{\rho}|\mathbf{u}|^2+\frac{a^2\rho^2}{\gamma \bar{\rho}}+\frac{\bar{\rho} c_v T^2}{\bar{T}}\right)dV,
\end{equation}
where the variables with $\Bar{\cdot}$ represent the base flow quantities.
For taking into account the energy norm, a similarity transform is applied to the resolvent operator. Therefore, for the singular value decomposition, we consider instead $\mathbf{H}_{\omega,\beta}\to W\mathbf{H}_{\omega,\beta} W^{-1}$. Here $W$ is a volume-weighted matrix that allows us to express the energy norm in terms of $L_2$ norm \citep{rolandi2024invitation}. 

In the present work, some cases are characterized by a linear dynamics that present eigenvalues with positive growth rate. For this reason, we use discounted resolvent analysis \citep{jovanovic2004modeling, jovanovic2005componentwise}, that considers a Laplace transform instead of a Fourier transform. This modification is equivalent to temporally damping the forcing and response by $e^{-\gamma t}$, which translates to considering the dynamic over finite timescales.  A zero temporal damping $\gamma=0$, when there are no unstable poles, corresponds to investigating the asymptotic (infinite-time) dynamics. Introducing $\gamma\neq 0$, we consider the dynamics over a finite-time horizon $t_\gamma=2\pi/\gamma$.  
To apply the discounting, the integration line of the Laplace transform is taken above all positive real parts (growth rate) of the eigenvalues, $\omega_r$. The discounted parameter $\gamma$ is thus introduced, which must satisfy $\gamma>\omega_r$. With discounting, the resolvent operator now reads  
\begin{equation}
    \mathbf{H}_{\omega,\beta}=[( \gamma-i\omega)\mathbf{I}-\mathcal{L}_\beta]^{-1}.
\end{equation}

The singular value decomposition of the resolvent operator $\mathbf{H}_{\omega,\beta}$ is approximated using the Krylov subspace projection method, with a subspace dimension for the reduced order problem set to $m=24$. Both the simulation of the base flow and resolvent analysis are performed within the compressible flow solver CharLES \citep{khalighi2011unstructured}, coupled with the PETSc and SLEPc libraries \citep{balay2020petsc,roman2016slepc} for the computation of the singular value decomposition. 

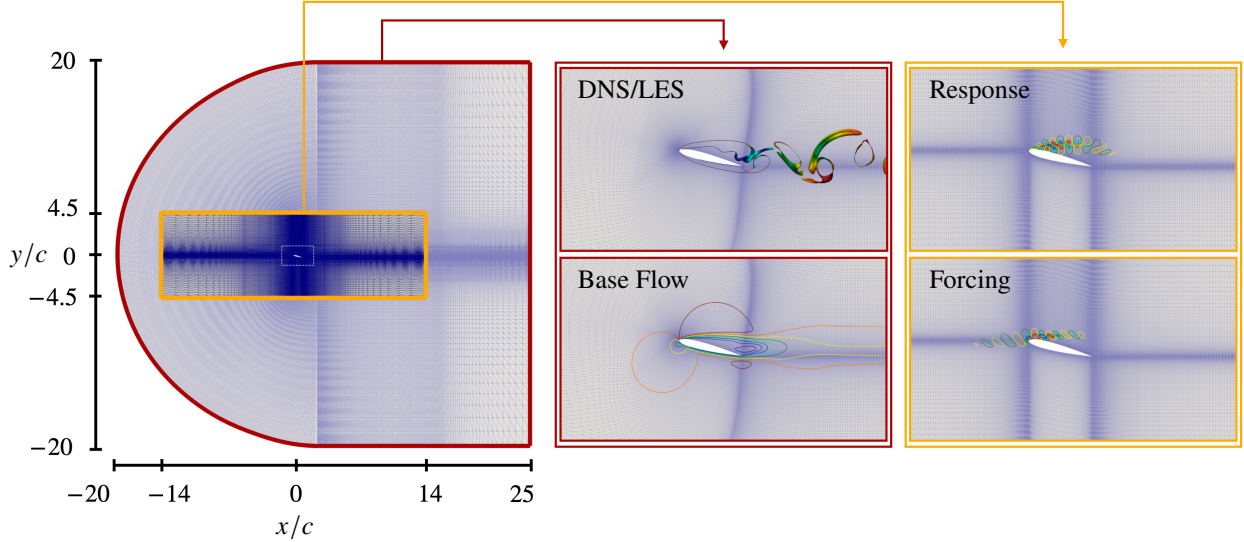
\begin{figure} 
\centering
\includegraphics[width=\textwidth]{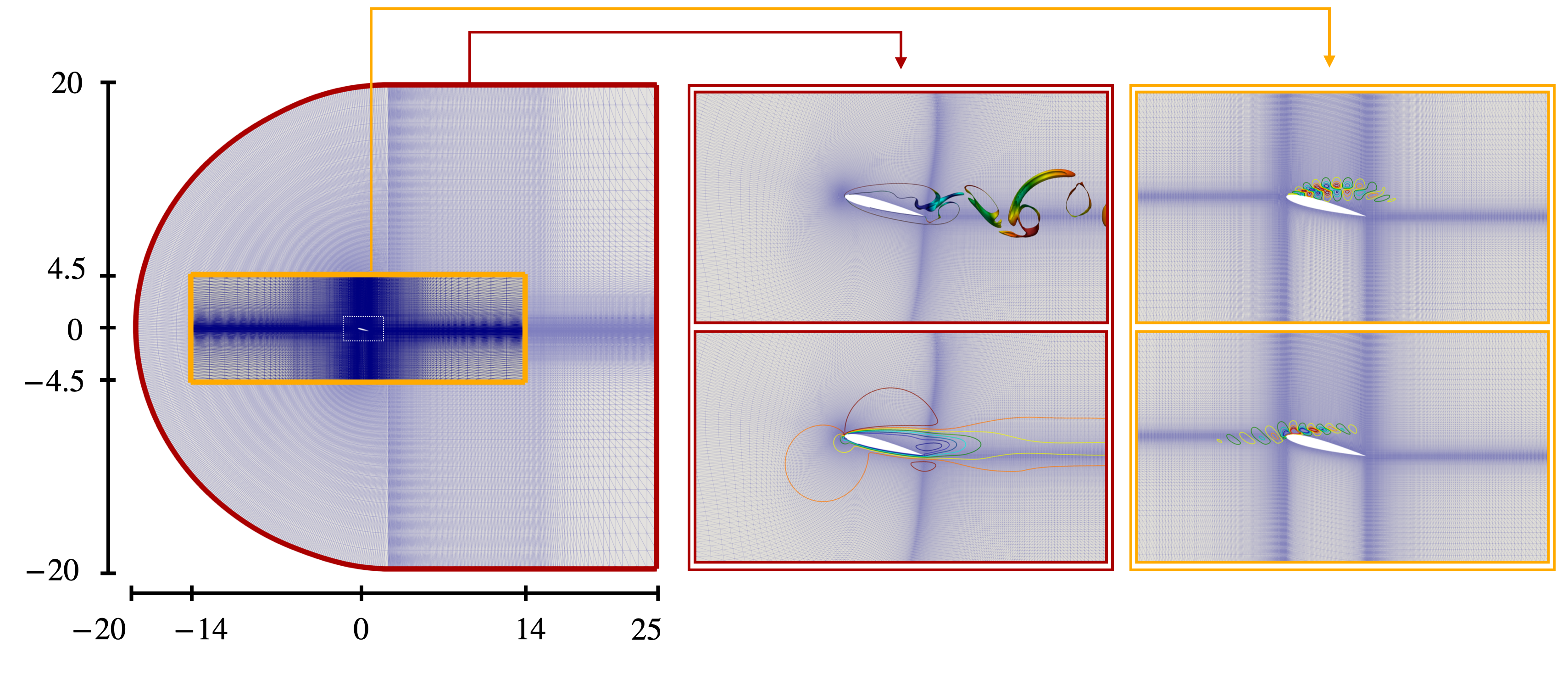}
\put(-255,165){{{DNS/LES}}}
\put(-122,165){{{Response}}}
\put(-122,93){{{Forcing}}}
\put(-255,93){{{Base Flow}}}
\put(-470,103){$y/c$}
\put(-369,0){$x/c$}
\caption{Computational setup used for the base flow computation and resolvent analysis. \label{fig:DetailsMesh}}
\end{figure}

\subsection{Computational setup}
We simulate the spanwise periodic flow around a two-dimensional NACA0012 airfoil at angles of attack $14^\circ$ and $9^\circ$ for a Reynolds number of $Re=U_\infty c/\nu=1000,\;2500,\;5000$ and $10000$ and a Mach number of $M_\infty=0.1$. Here, $U_\infty$ indicates the freestream velocity, $c$ the chord of the airfoil, and $\nu$ the kinematic viscosity. At these Reynolds numbers, the flow is three-dimensional so a spanwise domain length $L_z$ of one chord is considered for the computational domain.  This choice is based on observations indicating that the flow transitions to a three-dimensional state with a spanwise wavelength of approximately $\lambda_z\approx c/3$ \citep{gupta2023two}. The base flows are simulated with direct numerical simulations for $Re=1000$ and $2500$ and wall-resolved large eddy simulations for $Re=5000$ and $10000$. For the wall-resolved large eddy simulation, we use the Vreman subgrid-scale model \citep{vreman2004eddy}. The boundary conditions are comprised of a no-slip adiabatic condition on the airfoil surface, uniform constant velocity, pressure, and temperature at the inlet and on far-field boundaries, a zero-pressure gradient at the outlet, and periodic boundary conditions on the lateral sides.

Details of the mesh used in this study are shown in Figure~\ref{fig:DetailsMesh}. The red-framed domain indicates the computational domain used for computing the base flow and is extended to $x/c\in [-20,25]$, $y/c\in [-20,20]$ and $z/c\in [0, 1]$. The origin is positioned at the leading edge of the airfoil. For the biglobal resolvent analysis, the time- and spanwise-averaged base flow is interpolated onto a smaller two-dimensional grid, the yellow-framed grid, whose extent is $(x/c,y/c)\in[-14,14]\times[-4.5,4.5]$. This reduction is possible because the domain and grid resolution requirements for computing resolvent modes differ significantly from those used in unsteady simulations \citep{rolandi2024invitation}. Specifically, the modal structures are concentrated near the airfoil, eliminating the necessity for an extended domain, which is instead essential for the base flow simulations. Additionally, the grid refinement near the airfoil focuses on both the downstream and upstream regions, as the forcing modes develop upstream due to the convective nature of the amplification mechanisms.

\begin{table}
  \begin{center}
\def~{\hphantom{0}}
  \begin{tabular}{cccc}
                Base flow mesh        &  Coarse$_{\text{BF}}$ &  Baseline$_{\text{BF}}$ &  Refined$_{\text{BF}}$ \\[3pt]
      $\bar{C}_D$ &   0.2277  &  0.2180  &  0.2197 \\
                           
    $\bar{C}_L$& 0.6879  &  0.6430 &   0.6474\\
     Resolvent analysis mesh & Baseline$_{\text{R}}$  &  Baseline$_{\text{R}}$ &   Refined$_{\text{R}}$\\
      $St^*=\text{arg}\max_{St}\sigma_1$& 2.81  &  2.72 &   2.83\\
  \end{tabular}
  \caption{Values of time-averaged drag ($\bar{C}_D$) and lift ($\bar{C}_L$) coefficients from the unsteady simulation together with frequency of maximum amplification from resolvent analysis for the different meshes tested.}
  \label{tab:gridRes}
  \end{center}
\end{table}

\begin{figure}
\centering
\includegraphics[width=\textwidth]{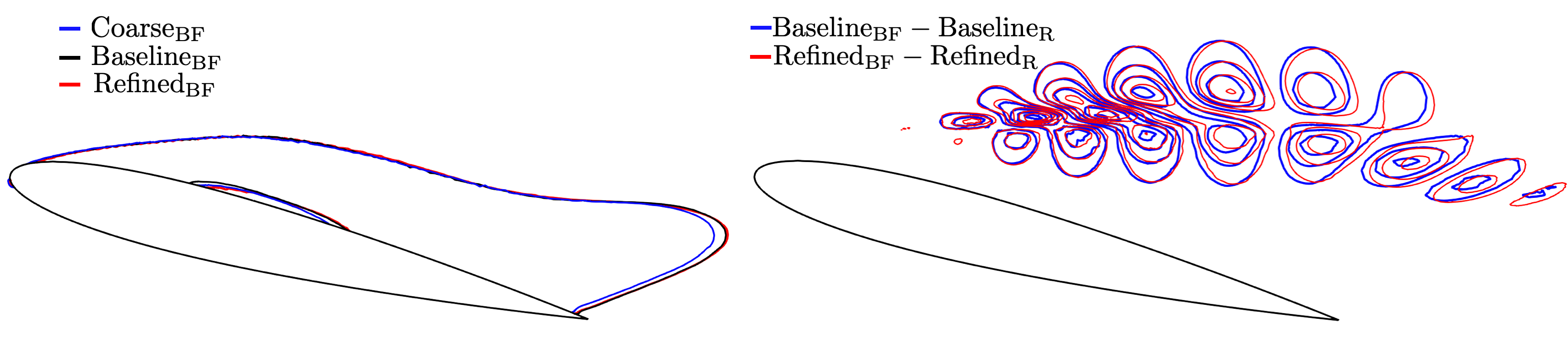}
\put(-455,108){\textit{(a)}}
\put(-245,108){\textit{(b)}}
\caption{ (a) Contour of time- and spanwise-averaged streamwise velocity $\bar{u}_x=0$ simulated with the three meshes for the unsteady simulation. (b) Contours of the streamwise velocity of the first response mode at frequency $\omega/2\pi=2.7$ for the two meshes, with base flows simulated with baseline and refined meshes for the unsteady calculations. \label{fig:GridConv} }
\end{figure}

 We performed a grid convergence study using three meshes for the unsteady simulation to compute the base flow: the coarse, baseline, and refined meshes. In addition to these meshes, two meshes were tested for the resolvent analysis: the baseline and refined meshes. In Table \ref{tab:gridRes}, we briefly report the results of the grid convergence study performed on the highest Reynolds number, $Re=10000$. The values of mean drag and lift coefficients from the unsteady simulation are shown, together with the frequency $St^*$ of maximum gain from the resolvent analysis. Resolvent analysis for the grid study was performed considering $\beta=0$ and $\gamma=1.25$. Overall, we can see convergence in the mean drag and lift coefficients and a relative difference between the results from the resolvent analysis within $5\%$. In Figure~\ref{fig:GridConv}, we present the $0$-contour of the streamwise velocity for the time- and spanwise-averaged base flow simulated using the three different meshes. Additionally, we display the streamwise velocity contours of the response mode at $\omega/2\pi=2.7$, computed using the baseline and refined meshes for the resolvent analysis. The base flows for these response modes were interpolated from the results of the unsteady simulations, computed with baseline and refined meshes, respectively. The results of both the base flow and resolvent analysis with the different meshes are in good agreement. The baseline meshes for both the unsteady simulation and resolvent analysis were used in this work.

\begin{figure}
\centering
\includegraphics[width=\textwidth]{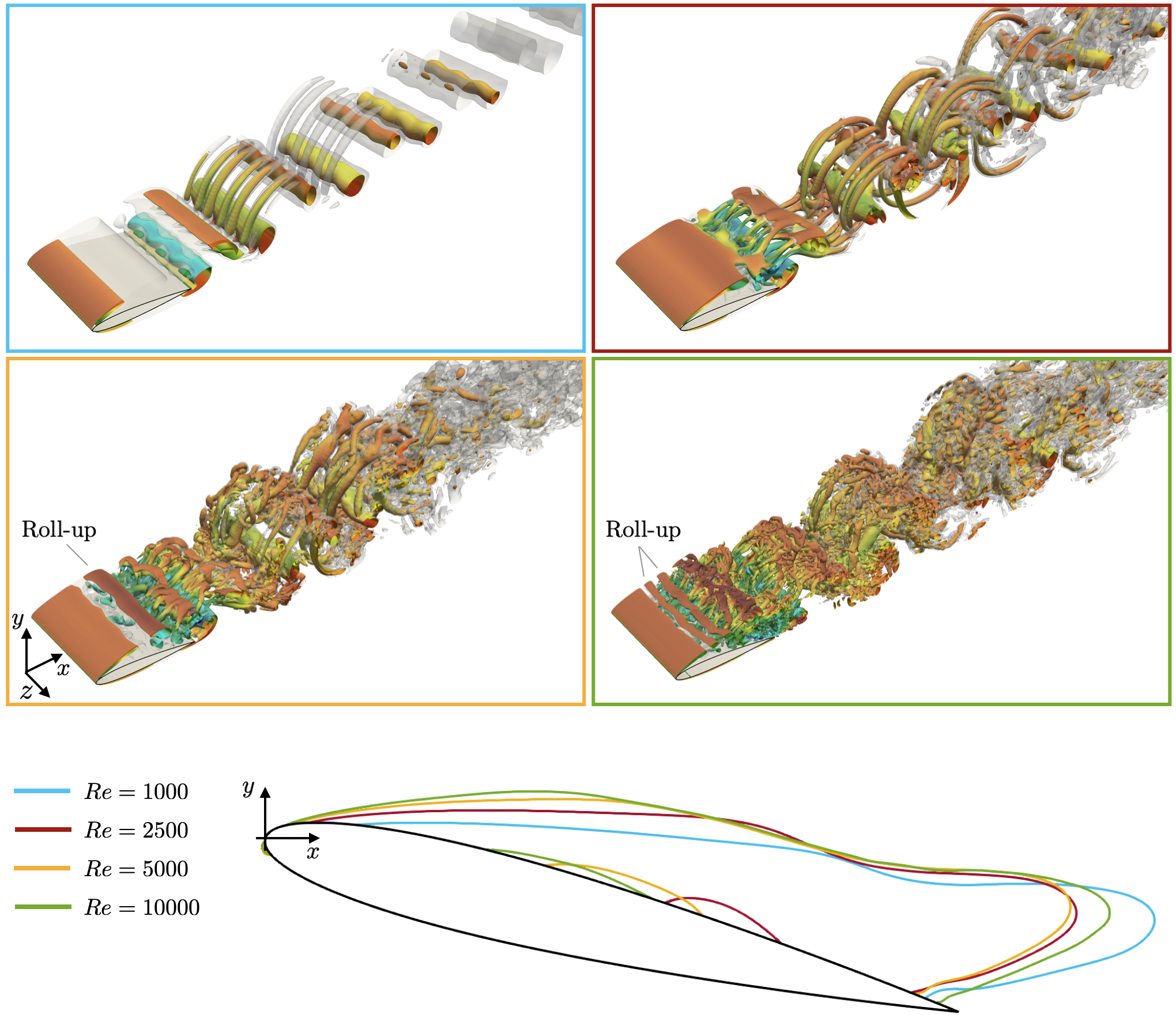}
\put(-460,397){\textit{(a)}}
\put(-460,255){\textit{(c)}}
\put(-225,397){\textit{(b)}}
\put(-225,255){\textit{(d)}}
\put(-355,105){\textit{(e)}}
\caption{Instantaneous flowfields around a NACA0012 wing at $\alpha=14^\circ$ and (a) $Re=1000$, (b $2500$, (c) $5000$ and (d) $10000$. Visualization with the isosurface of Q-criterion $Q=0.05$ colored by streamwise velocity superposed on translucent isosurface of Q-criterion $Q=0.005$. (e) Contour of time- and spanwise-averaged streamwise velocity $\bar{u}_x=0$.  \label{fig:baseflow} }
\end{figure}
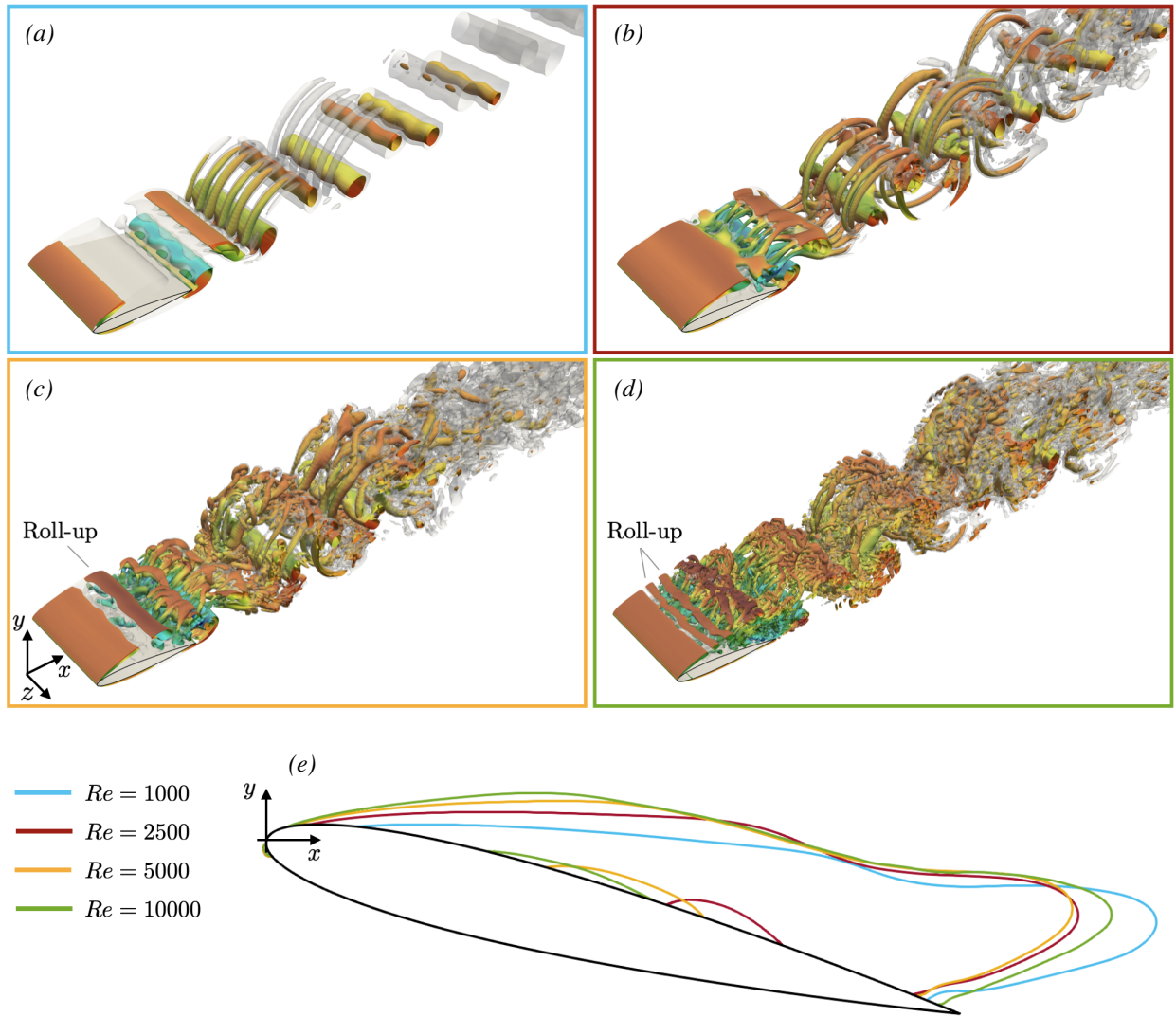

\section{Reynolds number effects on the unsteady and base flows}\label{sec:BaseFlow}
The flow fields around a NACA0012 airfoil at an angle of attack of $14^\circ$ for the considered Reynolds numbers are shown in Figure~\ref{fig:baseflow}.a-d. At these Reynolds numbers, the flow is unsteady and exhibits the characteristic von Kármán vortex shedding in the wake region. The flow around the airfoil at $\alpha=14^\circ$ undergoes a transition from steady to periodic state at approximately $Re\approx 380$ through a Hopf bifurcation \citep{victoria2022stability}. As the Reynolds number increases further, the periodic flow transitions to three-dimensional dynamics through a period-doubling bifurcation, also known as the Mode C instability  \citep{sheard2005subharmonic,meneghini2011wake,rolandi2021stability}. The Reynolds number at which the flow becomes three-dimensional at this angle of attack is between $Re=750$ and $1000$ \citep{gupta2023two}. This transition is characterized by the emergence of a subharmonic component of the vortex shedding, effectively doubling the flow periodicity. Notably, period-doubling bifurcations are also associated with the onset of chaos in fluid flows \citep{pulliam1993transition}, through the so-called period-doubling cascade.

The three-dimensional flow resulting from this transition features spanwise structures that develop across the airfoil. The Mode C instability develops in the stretched region between two consecutive vortices, called the braid region, and results in the formation of elongated streamwise vortices. At $Re=1000$, the spanwise wavelength of the three-dimensional structures is approximately $\lambda_z\approx c/3$, consistent with previous studies \citep{gupta2023two}, where the stability of the periodic solution was examined through Floquet analysis. As the Reynolds number increases further, the shear layer separating from the leading edge becomes unstable. The shear layer rolls up closer to the leading edge, forming two-dimensional spanwise vortical structures, that are related to the Kelvin-Helmholtz instability. The characteristic length of the vortical spanwise elongated structures decreases with the Reynold numbers (see Figure~\ref{fig:baseflow}.c-d). Finally, these structures are convected downstream, where they break down into smaller, three-dimensional structures, contributing to the increasing level of turbulence in the flow.

The corresponding zero streamwise velocity contours of the time- and span-averaged base flows are visualized in Figure~\ref{fig:baseflow}.e. At $Re=1000$, the base flow features a single recirculation region, which is elongated compared to those at higher Reynolds numbers. When the Reynolds number is increased to $Re=2500$, the recirculation region shortens before being stretched again at higher Reynolds numbers. For $Re\geq2500$, the base flow exhibits a 
secondary recirculation region on the airfoil's suction side, which shifts upstream as the Reynolds number increases, while thinning. The secondary recirculation is caused by the increasing reversed flow and the adverse pressure gradient forming in the separated region on the suction side. The interaction between the primary and the secondary recirculation regions has been seen to contribute to turbulent transition \citep{gao2019wall}. Additionally, the separation point of the primary recirculation region shifts upstream with the increasing Reynolds number \citep{counsil2013low,brunner2021study}.
This behavior is due to a higher adverse pressure gradient near the leading edge of the airfoil due to the thinning of the boundary layer over the airfoil upstream of the separation point with increasing Reynolds number.

\begin{figure}
\centering
\includegraphics[width=\textwidth]{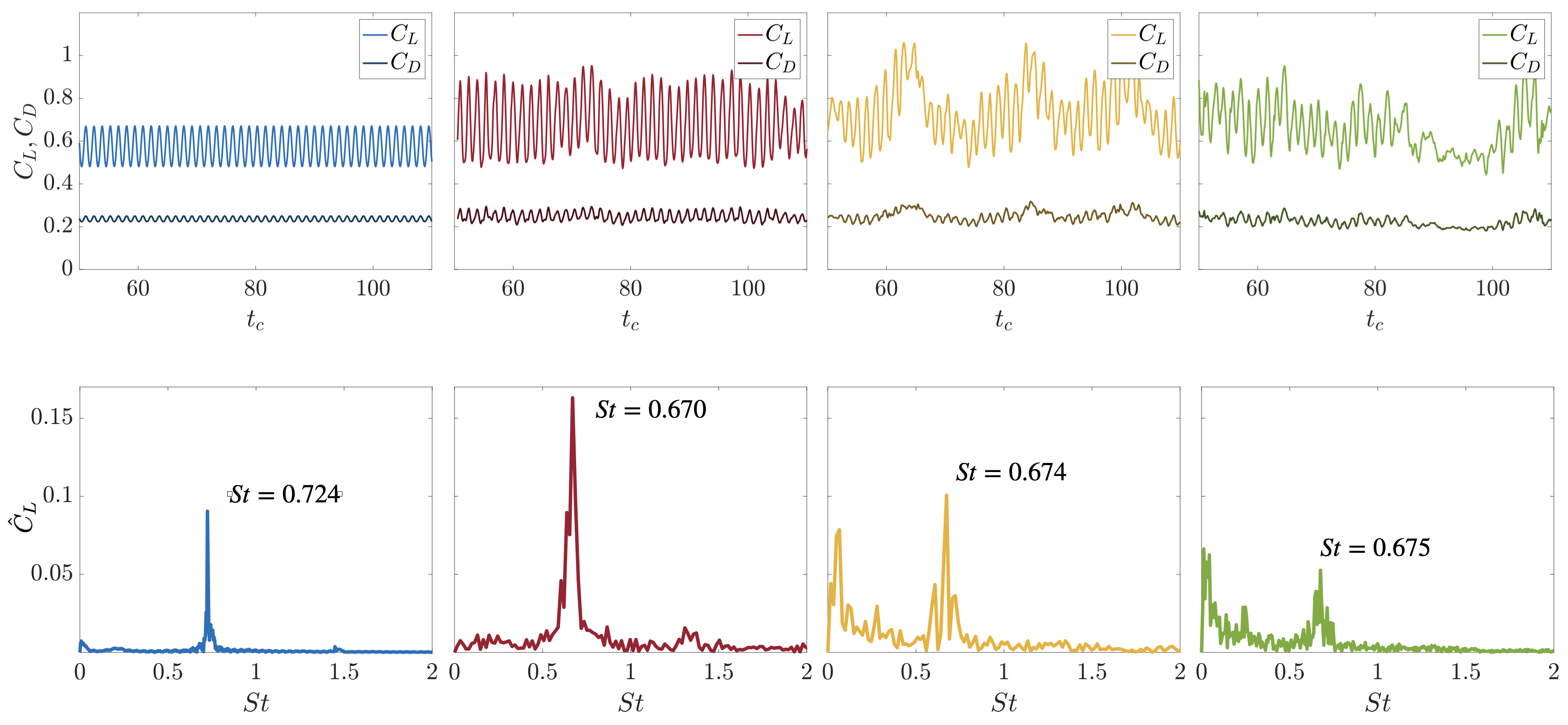}
\put(-448,220){\textit{(a)}}
\put(-332,220){\textit{(b)}}
\put(-222,220){\textit{(c)}}
\put(-112,220){\textit{(d)}}
\caption{Lift coefficient $C_L$, drag coefficient $C_D$ and lift spectra $\hat{C}_L$ at (a) $Re=1000$, (b) $2500$, (c) $5000$ and (d) $10000$. \label{fig:Strouhalbaseflow} }
\end{figure}
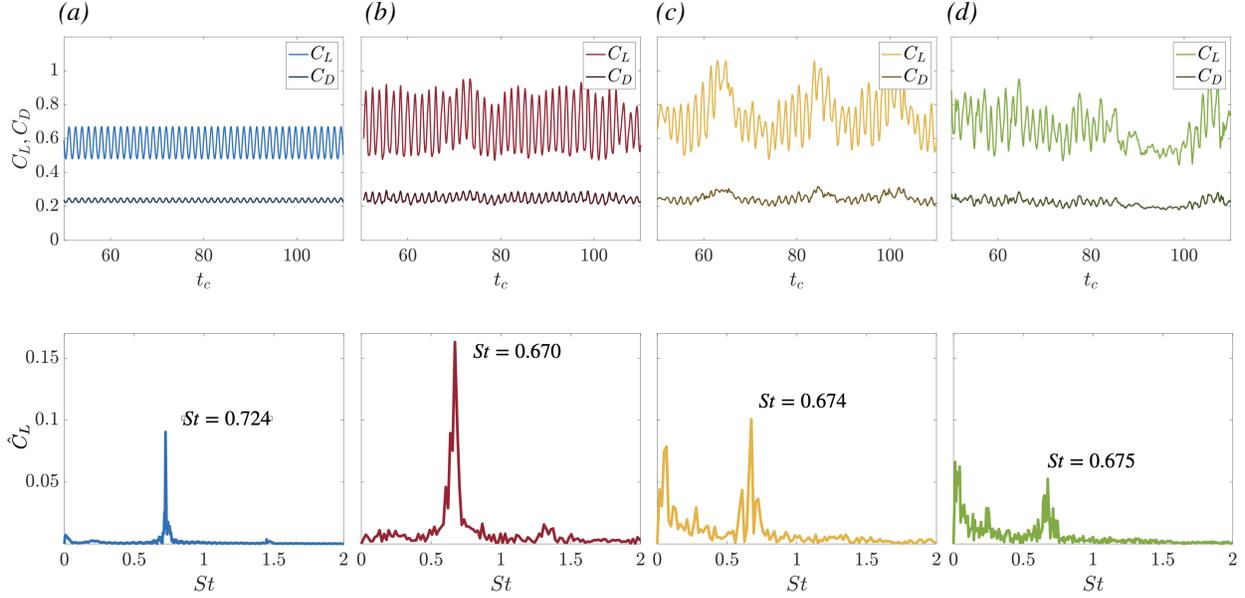

We report the time trace of the drag and lift coefficients in Figure~\ref{fig:Strouhalbaseflow}, together with the frequency spectra of the lift coefficients in terms of Strouhal number $St=c f/U_\infty$, where $f$ indicates the frequency. For $Re=1000$, we observe a clear peak at $St= 0.72$, due to the periodicity of the flow at this Reynolds number, while at $Re=2500$ the oscillations have higher amplitude and the peak is at a lower frequency of $St=0.67$. Increasing the Reynolds number, we observe a broader frequency spectrum and the emergence of slower dynamics, visible in the $C_L$ and $C_D$ variations. The highest frequency peak for both $Re=5000$ and $10000$ occurs at $St\approx0.67$, as the $Re=2500$ case. However, the amplitudes of the spectral peak reduce with the Reynolds number, consistent with the increasing irregularity.

\begin{figure}
\centering
\includegraphics[width=\textwidth]{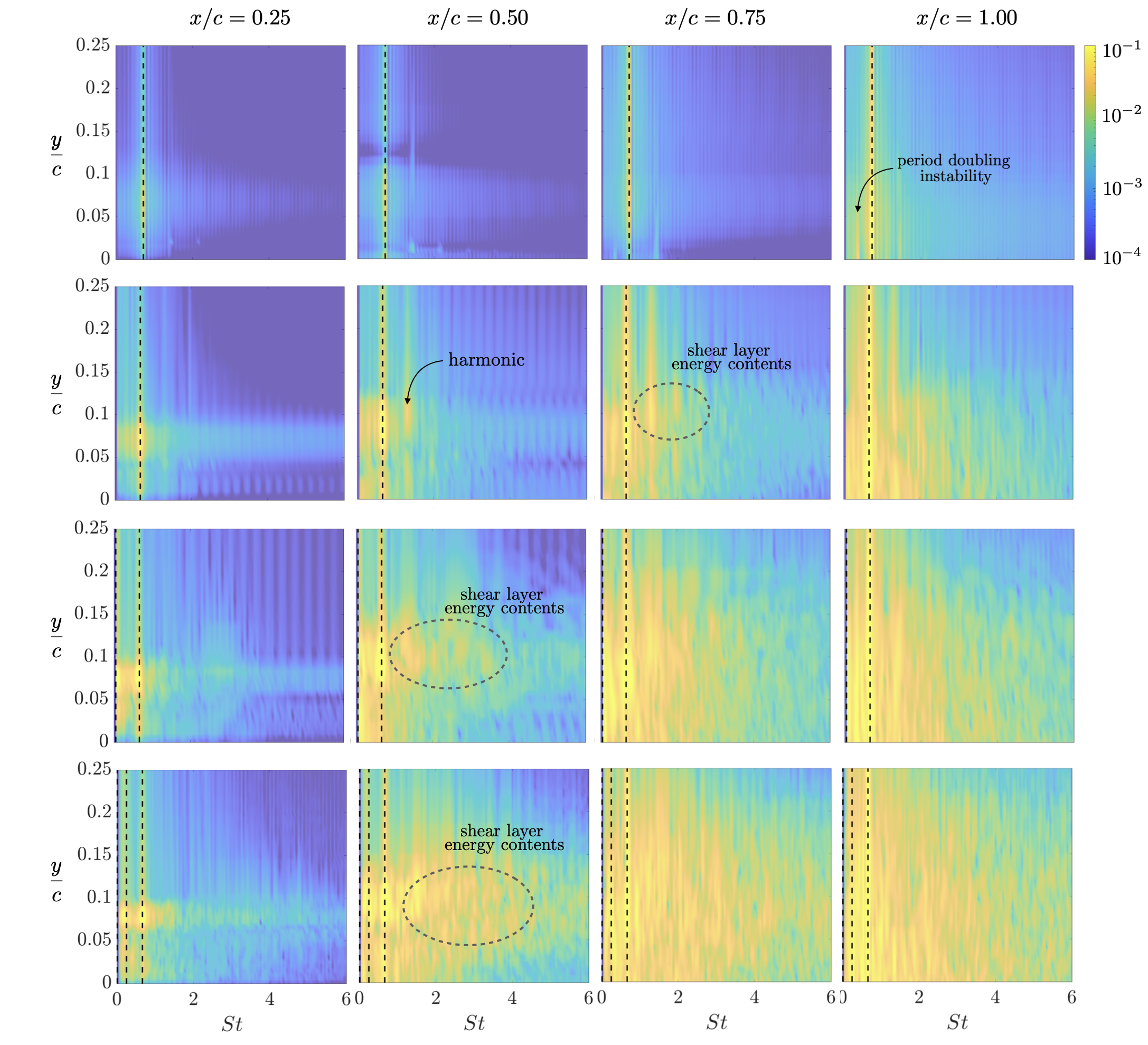}
\put(-463,362){\rotatebox[origin=c]{90}{$Re=1000$}}
\put(-463,263){\rotatebox[origin=c]{90}{$Re=2500$}}
\put(-463,165){\rotatebox[origin=c]{90}{$Re=5000$}}
\put(-463,70){\rotatebox[origin=c]{90}{$Re=10000$}}
\caption{Contours of energy spectra at $Re=1000$, $2500$, $5000$ and $10000$. Contours are shown at different streamwise locations $x/c$ along $y/c\in[0,0.25]$. Black dashed lines indicate the dominant frequency peaks associated with the lift coefficient, see Figure~\ref{fig:Strouhalbaseflow}. \label{fig:ProbeBOX} }
\end{figure}
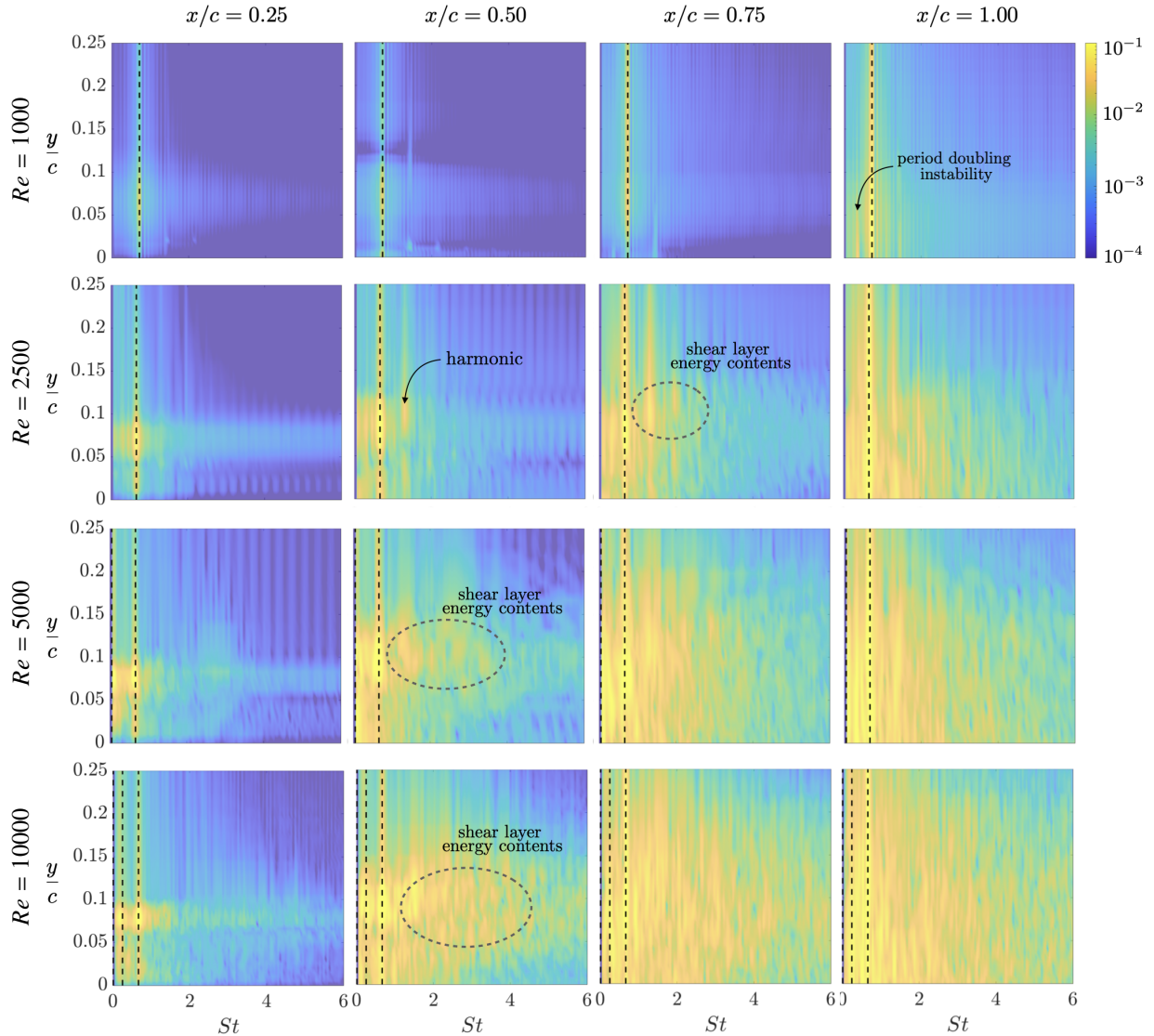

The lift coefficient signal effectively captures the dynamics of large vortex shedding in the wake, as demonstrated by the lift spectra in Figure~\ref{fig:Strouhalbaseflow}. However, our interest also lies in the dynamics of the separated shear layer, which manifests as the leading edge vortex roll-up seen in the instantaneous flow field in Figure~\ref{fig:baseflow}. To investigate the shear layer dynamics, we compute the energy spectra from the shear layer at various streamwise positions: $x/c = 0.25, 0.5, 0.75$ and $1$, along $0 < y/c < 0.25$, as shown in Figure~\ref{fig:ProbeBOX}. At $Re = 1000$, the flow is periodic, with the energy spectra at each position showing a dominant peak corresponding to vortex shedding. At $x/c = 1$, the sub-harmonic becomes significant, indicating a period-doubling instability of the NACA0012 periodic shedding that leads the transition to three-dimensional flow \citep{sheard2005evolution, meneghini2011wake, rolandi2021stability, gupta2023two}. At $Re = 2500$, the harmonic components become more prominent, and at even higher Reynolds numbers, the spectrum broadens significantly, especially as the streamwise position approaches the wake. Also for the higher Reynolds number cases, we observe the presence of subharmonics, which, in this case, correspond to vortex pairing within the shear layer. For $Re \geq 2500$, the highest energy amplitudes at $x/c = 0.25$ are concentrated at a specific cross-stream location, approximately $y/c \approx 0.75$. This identifies a broad region of high-frequency, high-amplitude unsteadiness associated with the shear layer dynamics at the subsequent streamwise positions.
 
\section{Resolvent analysis}\label{sec:ResolventRes}

In this section, we present the results of the resolvent analysis, organized into three parts. First, we examine the eigenvalues of the linear operators to establish an appropriate range for the discount parameter $\gamma$. This enables us to study the response mode structures and the variations in energy gain as a function of 
$\gamma$. Second, we investigate the influence of the spanwise wavenumber $\beta$ on the system. Finally, we examine how modal structures and energy gain vary with the Reynolds number across a range of frequencies.

\begin{figure}
\centering
\includegraphics[width=\textwidth]{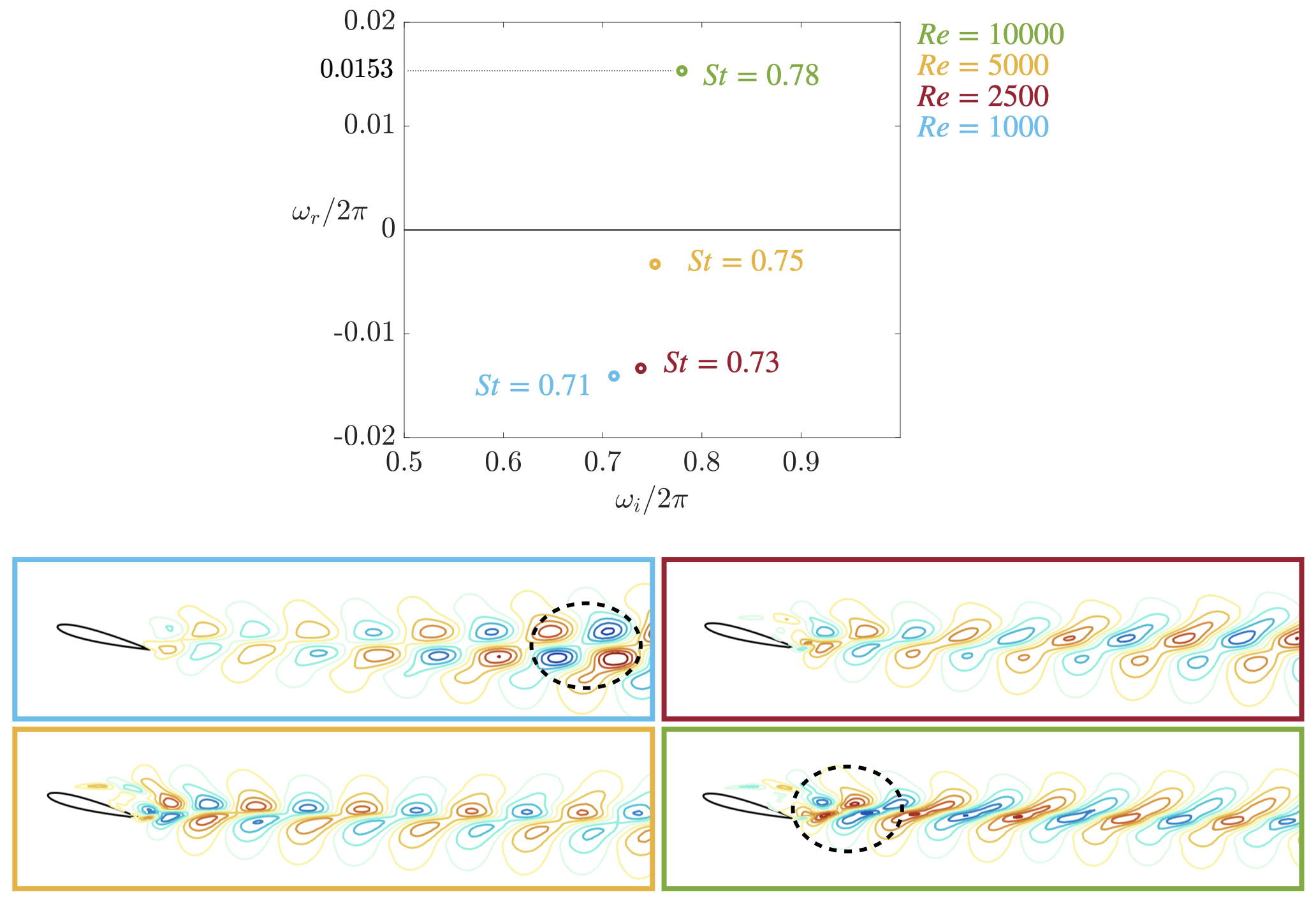}
\put(-370,318){\textit{(a)}}
\put(-465,130){\textit{(b)}}
\caption{(a)  The eigenvalues with the largest real components and (b) corresponding eigenvectors shown by contours of real part streamwise velocity, with maximum modal structure amplitude circles in dashed line. \label{fig:Stability} }
\end{figure}
\subsection{Eigenvalue decomposition of the linear operator}
Let us examine the eigenvalues of the linearized NS operator at the different Reynolds numbers by solving the eigenvalue problem 
\begin{equation}
    \mathcal{L}\mathbf{\phi}=\tilde{\omega}\mathbf{\phi},
\end{equation}
where $\tilde{\omega}=\omega_r+i\omega_i$ is the complex eigenvalue and $\mathbf{\phi}$ is the corresponding eigenvector. Here, we consider $\mathcal{L}=\mathcal{L}_{\beta=0}$, while the effect of spanwise wavenumber $\beta> 0$ will be explored in Sect. \ref{sec:BetaEffects}.  This allows us to consider appropriate ranges for the discount parameter depending on whether we consider two-dimensional $\beta=0$ or three-dimensional $\beta\neq0$ perturbations.

In Figure~\ref{fig:Stability}.a, the eigenvalue with the largest real part for each $Re$ is shown in the complex plane. Both the growth rate and frequency increase with increasing Reynolds number.  The linear dynamics holds an eigenvalue with a positive real part only at $Re=10000$, with $\omega_r/2\pi=0.0153$. For the lower Reynolds number cases, all the eigenvalues hold negative real parts, despite the unsteady nature of the nonlinear dynamics. In this regard, it should be considered that the eigenvalue analysis depends on the choice of the base flow and that there are conditions for the validity of mean flow stability analysis \citep{beneddine2016conditions}. Nevertheless, from the result shown in Figure~\ref{fig:Stability}.a, we observe a monotonic increase of the $\omega_r$ with respect to the Reynolds number, particularly a linear increase for $Re\geq2500$. We can then infer that the eigenvalue based on the time-averaged base flow crosses the imaginary axis at $Re\approx 5900$. 
The corresponding eigenvectors are shown in Figure~\ref{fig:Stability}.b. At $Re=1000$ the modal structure is mainly concentrated in the wake region, while at higher Reynolds numbers they also exhibit structures in the shear layer region. As the Reynolds number increases, the position of maximum modal structure amplitude, circled by the dashed line, shifts from the far wake (at $Re=1000$) to the near wake (at $Re=10000$).

\subsection{Temporal discounting}\label{sec:FiniteHorizon}
Once the eigenvalues are found, the discount parameter for the resolvent analysis is chosen such that $\gamma/2\pi>\max\{\omega_r/2\pi\,0\}$. This allows us to consider the overall forced dynamics within a timescale smaller than the timescale associated with $\max\{\omega_r\}$. Larger timescales should not be considered, because the implication of having a positive real part of $\omega$ would make the response seemingly unbounded at $t\to \infty$, masking the effect of forcing. 

\begin{figure} 
\centering
\includegraphics[width=\textwidth]{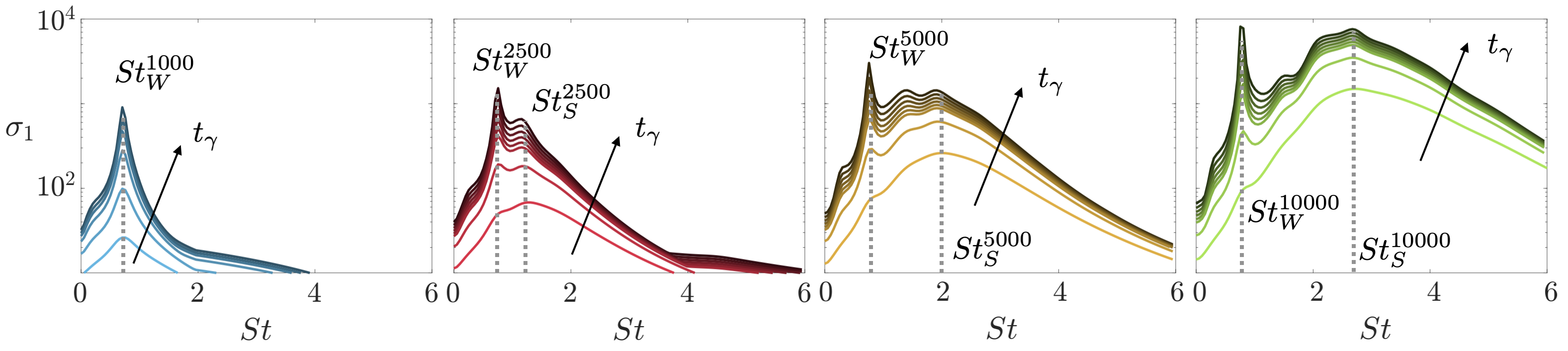}
\put(-418,105){{$Re=1000$}}
\put(-305,105){{$Re=2500$}}
\put(-192,105){{$Re=5000$}}
\put(-85,105){{$Re=10000$}}
\caption{Variation of the first sigular value $\sigma_1$ over the frequency for the different finite-time horizon $t_\gamma \in [5;41.6]$ at $Re=1000$, $2500$, $5000$ and $10000$. Dashed gay lines indicate the frequencies of maximum gain at short and long timescales.  \label{fig:Gain1_tGamma}}
\end{figure}

Here, a value of $\gamma/2\pi >0.0153$ for $Re=10000$ corresponds to dynamics within a finite time horizon $t_\gamma<65.3$. The effect of discounting on the energy amplification is shown in Figure~\ref{fig:Gain1_tGamma}.  The variations of the first singular value over frequency are plotted for $\gamma=\{0.15,0.20,0.25,0.30,0.40,0.625,1.25\}$ corresponding to $5\leq t_\gamma=2\pi/\gamma\leq 41.9$. The frequency is considered in terms of Strouhal number $St=c\omega/U_\infty2\pi$.
\begin{figure} 
\centering
\includegraphics[width=0.7\textwidth]{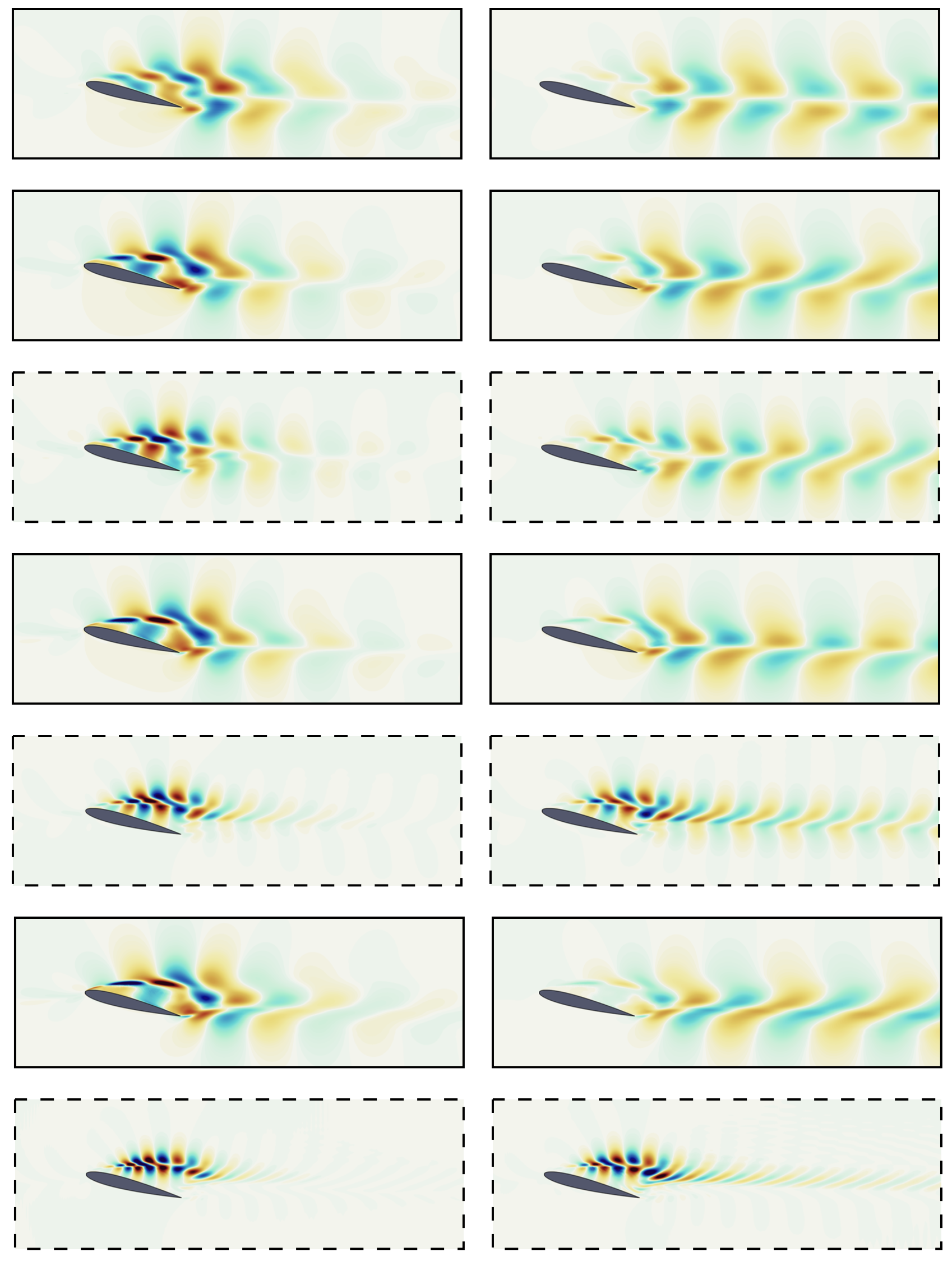}
\put(-260,440){$t_{\gamma,\text{short}}$}
\put(-100,440){$t_{\gamma,\text{long}}$}
\put(-340,402){\rotatebox[origin=c]{90}{$Re=1000$}}
\put(-340,305){\rotatebox[origin=c]{90}{$Re=2500$}}
\put(-340,180){\rotatebox[origin=c]{90}{$Re=5000$}}
\put(-340,60){\rotatebox[origin=c]{90}{$Re=10000$}}
\put(-320,420){$St_W^{1000}$}
\put(-320,356){$St_W^{2500}$}
\put(-320,293){$St_S^{2500}$}
\put(-320,230){$St_W^{5000}$}
\put(-320,167){$St_S^{5000}$}
\put(-320,105){$St_W^{10000}$}
\put(-320,42){$St_S^{10000}$}
\caption{Streamwise velocity component of the first response mode at the frequencies of maximum gain at short and long timescales. (\full) line frame indicates the mode at the lower frequency peak, $St_W$, and (\dashed) line frame indicates the mode at the higher frequency peak, $St_S$.    \label{fig:DiscoiuntRespModes}}
\end{figure}
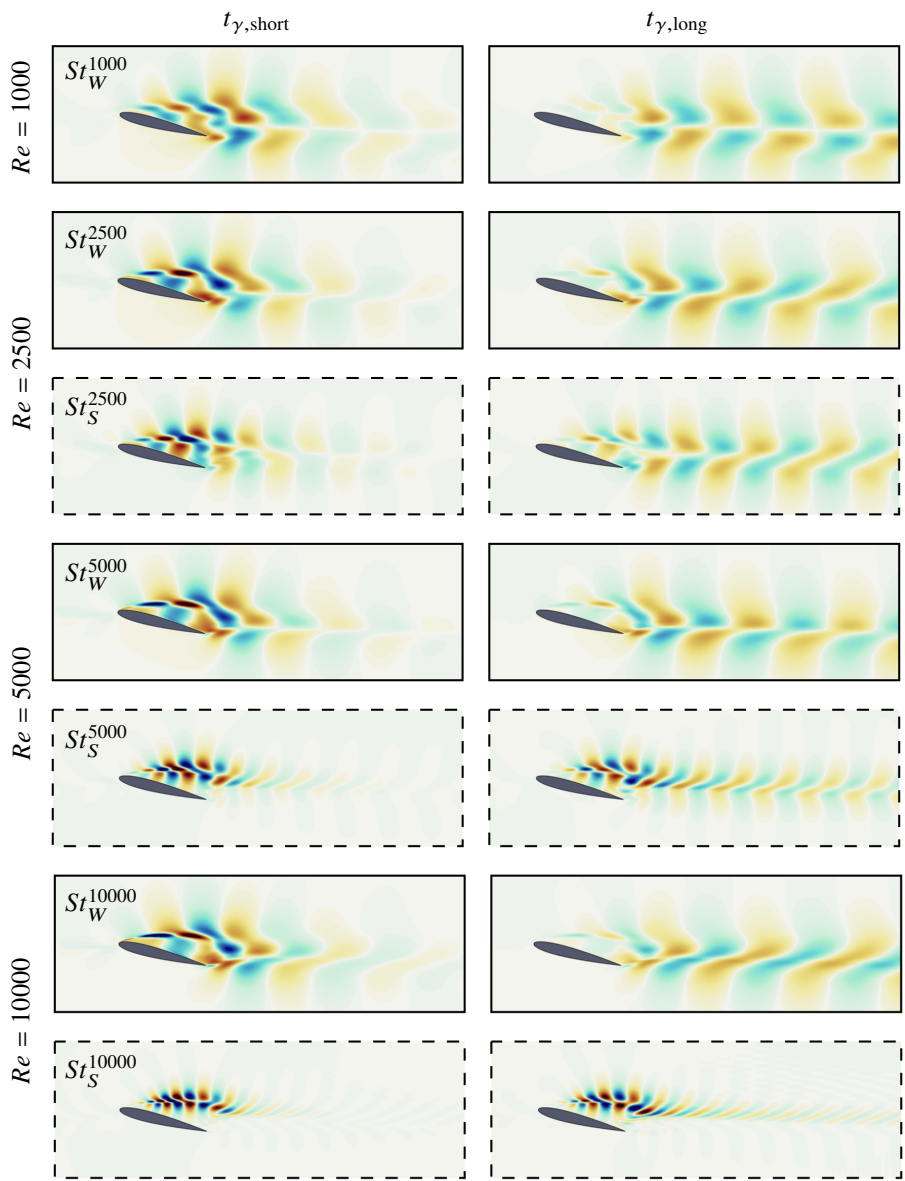
Considering this range of $t_\gamma$, in what follows we will refer to $t_\gamma=5$ as the short timescale and $t_\gamma=41.9$ as the long timescale. At $\Re=1000$, only one peak emerges as $t_\gamma$ varies, while for $Re\geq2500$ two distinct peaks are observed, as indicated by the dashed lines in Figure~\ref{fig:Gain1_tGamma}. The first peak appears at a high frequency on the short timescale, while on the long timescale, another peak at a lower frequency dominates. For all Reynolds numbers, the long timescale peak occurs at a frequency corresponding to the eigenvalue with the maximum real part. This is because, with lower $\gamma$ (higher $t_\gamma$), the Laplace integration is closer to such an eigenvalue, and the norm of the resolvent operator increases at that frequency.

In Figure~\ref{fig:DiscoiuntRespModes}, we show, for the considered Reynolds numbers, the streamwise velocity components of the first response mode at short and long timescale and at the frequencies $St_W$ and $St_S$, indicated in Figure~\ref{fig:Gain1_tGamma}. Frequencies $St_W$ and $St_S$ correspond to the short and long timescale peaks, respectively. At the lower frequencies, $St_W$, we observe that the structures emerge in the wake and highlight the coupling between the leading and trailing edge, we will therefore refer to this mode as the wake mode. We also note that the modal structures at the lowest Reynold number are similar to the structures revealed from non-modal stability analysis of low-Reynolds number flow \citep{he2017linear} when increasing the time horizon. On the other hand, at the higher frequency peak, $St_S$, the structures are present in the shear layer region detaching from the leading edge.  We therefore refer to this mode as the shear layer mode. For both the wake and shear layer modes, we observe that the structures at $t_{\gamma,\text{short}}$ are closer to the airfoil, while they develop downstream when increasing $t_\gamma $. This corresponds to the fact that the perturbation has more time to grow, and translates into the higher energy gain shown in Figure~\ref{fig:Gain1_tGamma}. Further discussion on the effects of discounting can be found in Appendix \ref{Appendix1}.

\begin{figure} 
\centering
\includegraphics[width=0.6\textwidth]{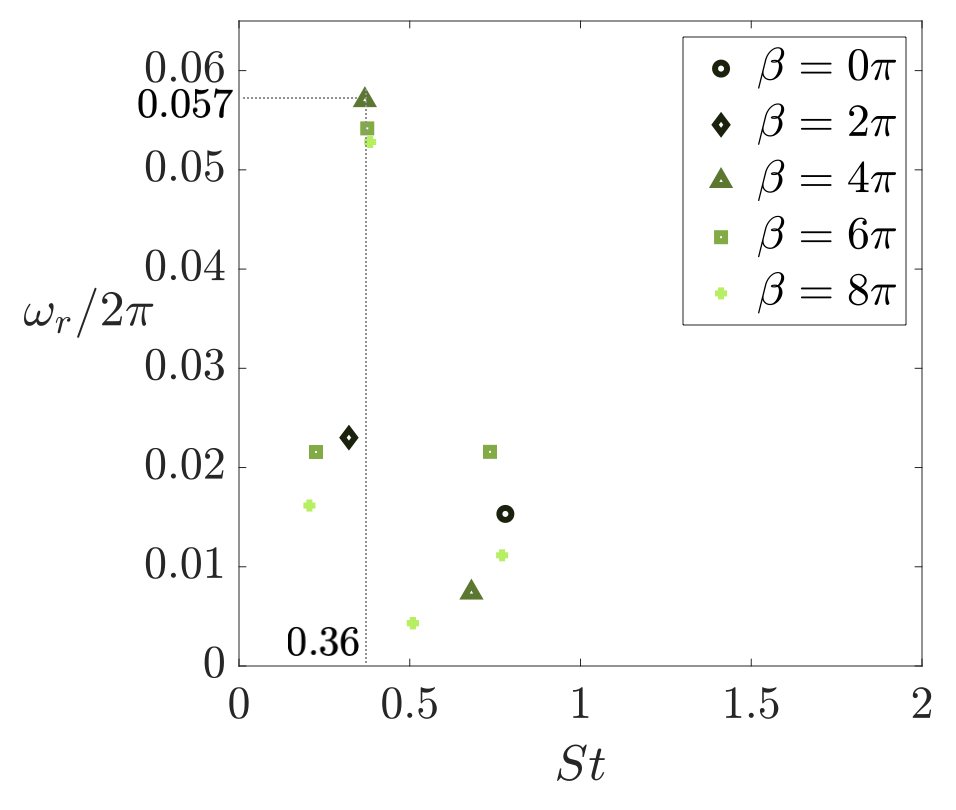}
\caption{The eigenvalues with the largest real components for different $\beta$ at $Re=10000$.   \label{fig:BetaEV}}
\end{figure}

\begin{figure} 
\centering
\includegraphics[width=\textwidth]{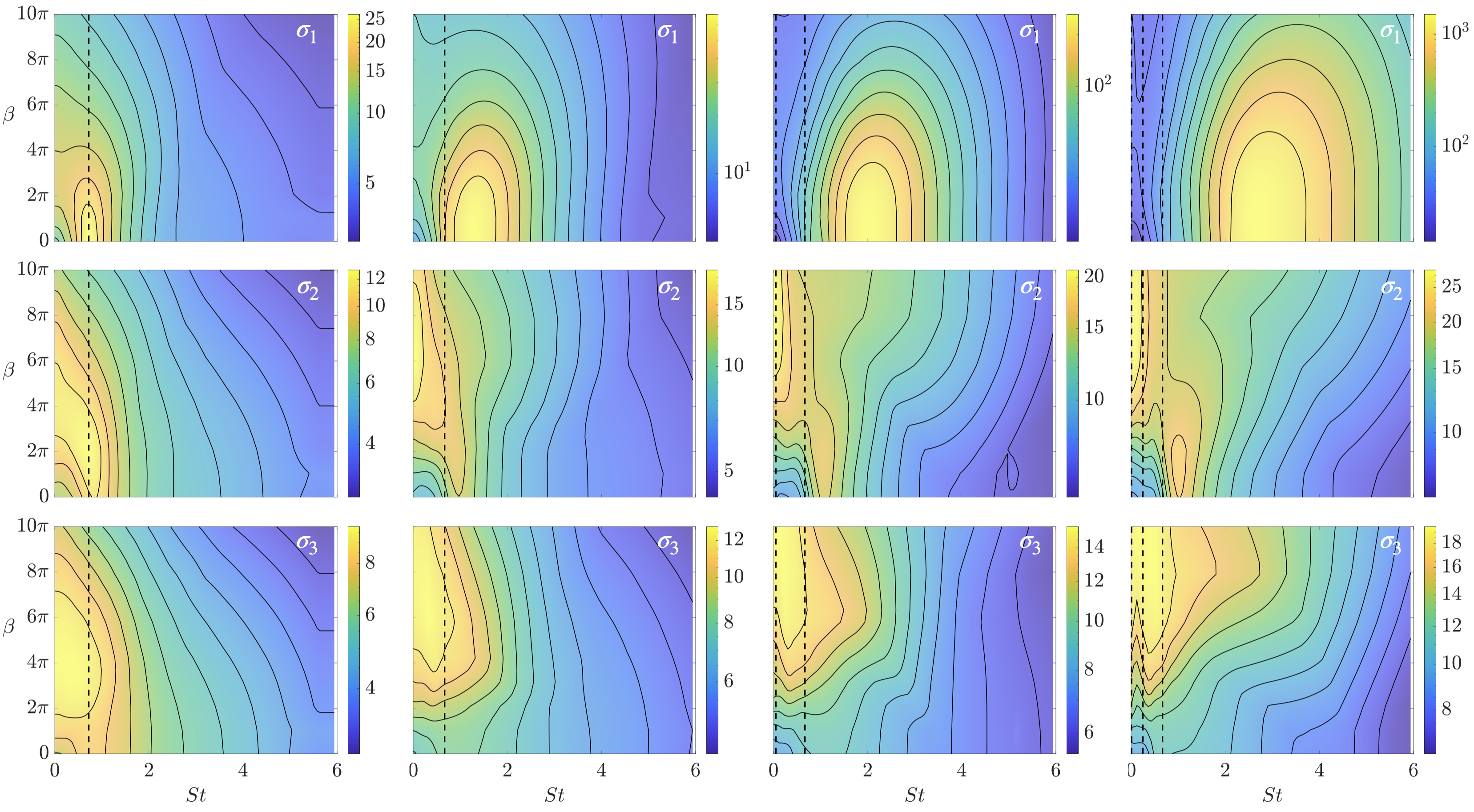}
\put(-430,260){$Re=1000$}
\put(-315,260){$Re=2500$}
\put(-200,260){$Re=5000$}
\put(-90,260){$Re=10000$}
\caption{Gain distributions of the first three singular values over the $\beta-St$ plane at $Re=1000$, $2500$, $5000$ and $10000$ at $t_\gamma=5$. Black dashed lines indicate the dominant frequency peaks associated with lift coefficients, see Figure~\ref{fig:Strouhalbaseflow}.   \label{fig:BetaSigmaAoA14_tshort}}
\end{figure}

\subsection{Spanwise wavenumber effects}\label{sec:BetaEffects}
\begin{figure} 
\centering
\includegraphics[width=\textwidth]{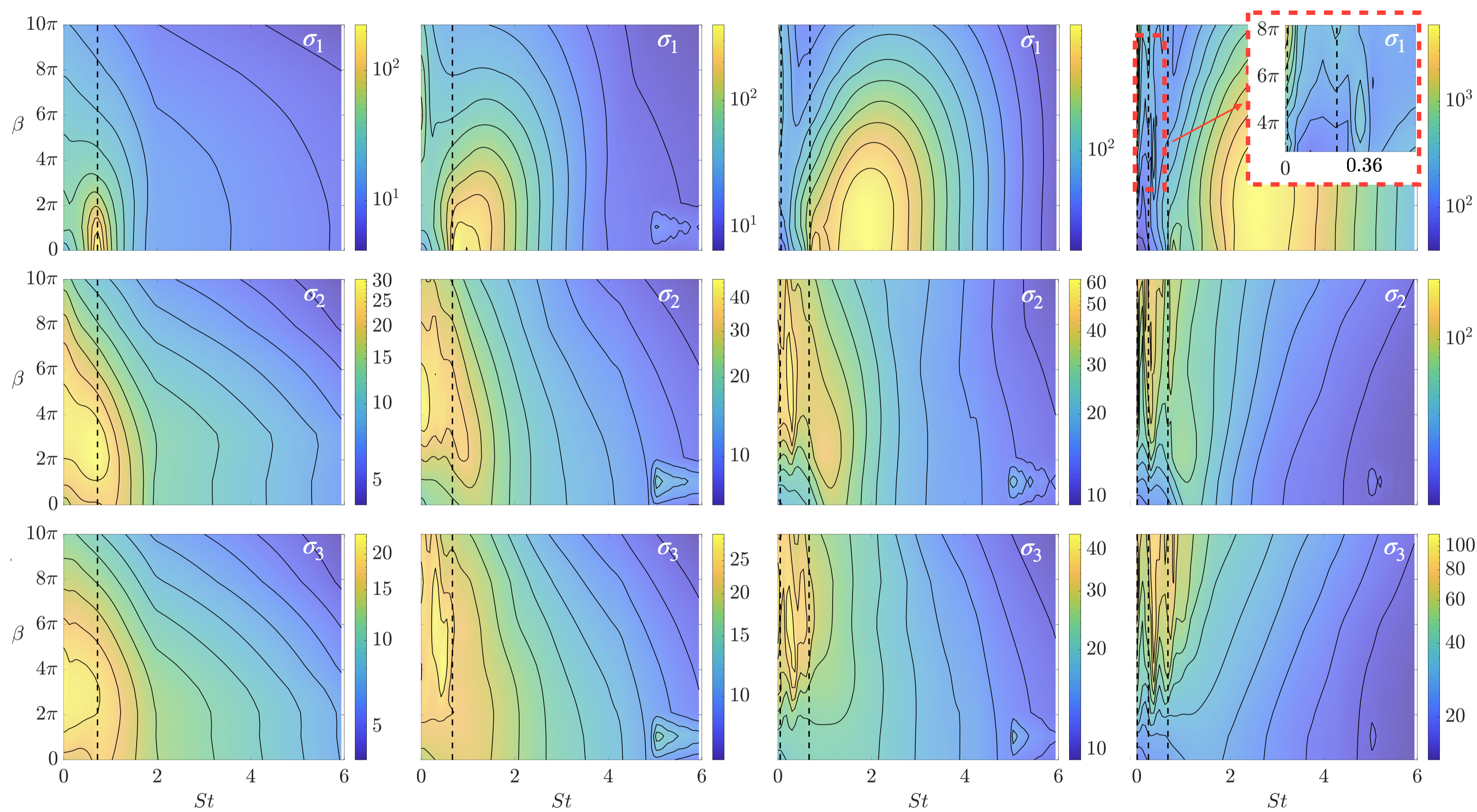}
\put(-430,260){$Re=1000$}
\put(-315,260){$Re=2500$}
\put(-200,260){$Re=5000$}
\put(-90,260){$Re=10000$}
\caption{Gain distributions of the first three singular values over the $\beta-St$ plane at $Re=1000$, $2500$, $5000$ and $10000$ at $t_\gamma=15$. Black dashed lines indicate the dominant frequency peaks associated with lift coefficients, see Figure~\ref{fig:Strouhalbaseflow}.   \label{fig:BetaSigmaAoA14_tlong}}
\end{figure}

In this subsection, we consider the effects of the spanwise wavenumber $\beta$. Firstly, we need to compute the eigenvalues of the linear operator $\mathcal{L}_\beta$ at varying $\beta$, as we performed for the $\beta=0$ case. In Figure~\ref{fig:BetaEV} the unstable eigenvalues are shown in the complex plane for $\beta\in[0;8\pi]$. The eigenvalue with the largest real part corresponds to $\beta=4\pi$ at $St=0.36$, and the real part decreases at the same frequency for increasing wavenumber. In this case, the value of the largest real part $\omega_r/2\pi=0.057$ suggests that we should consider dynamics within a timescale of $t_\gamma=17.54$, thus shorter compared to the $\beta=0$ investigated in the previous subsection. 
The effects of spanwise wavenumber $\beta$ at short, $t_\gamma=5$, and medium, $t_\gamma=15$, timescales are shown in Figures \ref{fig:BetaSigmaAoA14_tshort} and  \ref{fig:BetaSigmaAoA14_tlong}, respectively. The gain distributions for the first three singular values ($\sigma_1,\;\sigma_2$ and $\sigma_3$) are shown over the $\beta-St$ plane for the different Reynolds numbers. The peaks of the spectral content of the lift coefficient, shown in Figure~\ref{fig:Strouhalbaseflow}, are also reported for comparison. 

For the short timescale, $t_\gamma=5$, we observe that the overall distributions of $\sigma_1,\sigma_2$ and $\sigma_3$ show some similarities across different Reynolds numbers. For $\sigma_1$, the maximum gain is achieved at $\beta=0$.  The singular values $\sigma_2$ and $\sigma_3$ are instead more sensitive to the spanwise variation. In particular, 
the variation of $\sigma_2$ and $\sigma_3$ across $\beta$ and $St$ are similar, with maximum values achieved at low frequency and spanwise wavenumbers that increase with the Reynolds number.
Overall, the second and third singular modes, even at short timescales, seem to reflect the development of smaller structures in the flow when increasing the Reynolds number, while the first singular value mostly reflects the two-dimensional dynamics.

The matter changes when we consider longer timescales, as shown in Figure~\ref{fig:BetaSigmaAoA14_tlong}. Increasing $t_\gamma$, several mode switchings are observed. These appear at low frequencies, particularly close to the characteristic frequencies at the highest Reynolds numbers and evident from a change in the gain variation over the frequency. This indicates that higher-order modes, which at short timescale reflect the relevance of finer spanwise structures, need more time to grow and overcome the energy of two-dimensional mechanisms that prevail at short timescale. At $Re=10000$, we show a zoomed-in view of the low $St$ and high $\beta$ parametric space, showing the emergence of local maximum at $St\approx 0.36$ and $\beta\approx 4\pi$, which reflects the large real part eigenvalue presented in Figure~\ref{fig:BetaEV}. 

Spanwise effects are thus seen to affect the low-frequency dynamics at long timescale. At higher frequencies, instead, the maximum gain remains close to $\beta=0$, and it is not affected by the timescale. This is because higher frequencies are linked to shear layer dynamics, which correspond to the quasi-two-dimensional roll-up of the shear layer separating from the leading edge, which is visible in Figure~\ref{fig:baseflow}.c-d.

In this section, we have analyzed the impact of the discount parameter, which accounts for dynamics over varying finite timescales, $t_\gamma$. Two dominant amplification mechanisms are revealed over the timescale. Over short timescales, the shear layer dynamics predominates, while at long timescales, the wake dynamics becomes the most energetic. Additionally, we have observed that spanwise wavenumber effects significantly influence the first singular modes only over longer timescales, whereas two-dimensional mechanisms dominate over short timescales. In agreement with the present results at $Re=1000$, previous studies at lower Reynolds numbers also report two-dimensional vortex shedding mechanism to be predominant \citep{he2017linear,nastro2023global}. By increasing the Reynolds number, this is true for the shear layer, while less energetic three-dimensional mechanism occur at lower frequencies.

\subsection{Reynolds number effects on energy gain, forcing and response modes at fixed \texorpdfstring{$\gamma$}{gamma}}\label{sec:ResultResolvent1}
In this section, we analyze the effect of the Reynolds number on the resolvent modes and gains at a fixed discount parameter. We choose the discount parameter to correspond to the short timescale $t_\gamma=5$ and show results at a spanwise wavenumber of $\beta=0$.
\begin{figure} 
\centering
\includegraphics[width=\textwidth]{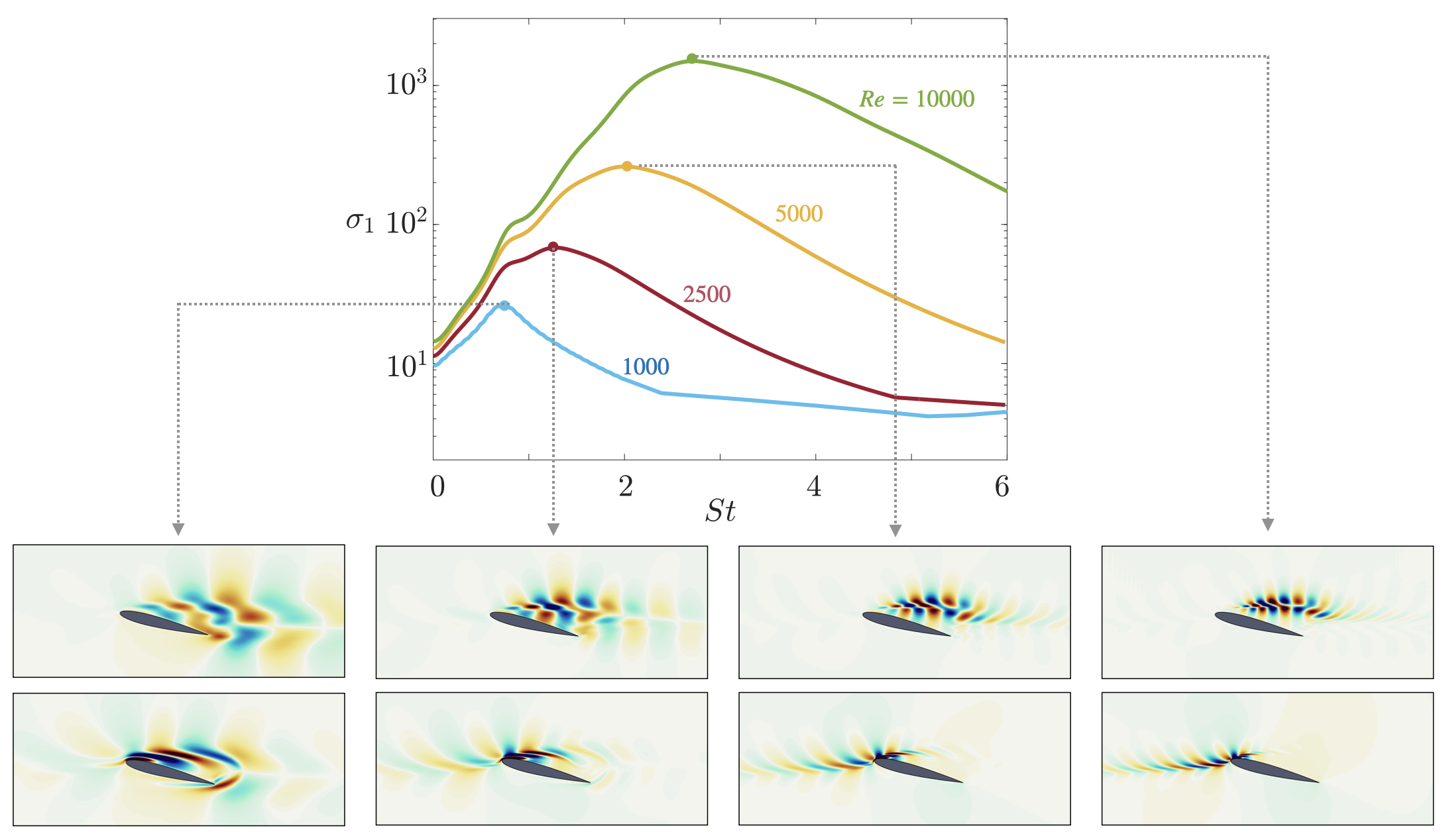}
\put(-460,85){$\hat{q}_{u_x}$}
\put(-460,35){$\hat{f}_{u_x}$}
\caption{Variation of the first singular value $\sigma_1$ with respect to the frequency for the different Reynolds numbers and response/forcing mode structures at the respective frequency of maximum gain.  \label{fig:SigmaModesAoA14}}
\end{figure}
 Figure~\ref{fig:SigmaModesAoA14} shows the variation of the first singular value over the frequency for different Reynolds numbers. The gain monotonically increases with increasing Reynolds number, and the maximum gain is achieved at higher frequencies, with the peak becoming less pronounced as the Reynolds number increases. It is important to note that the peak at $Re=1000$ corresponds to a wake dynamics influenced by the most unstable eigenvalue, and the ``bumps" at $St\approx0.7$ for $Re\geq2500$ are the short-time effect of the eigenvalues with the largest real part, as pointed out in Sect.~\ref{sec:FiniteHorizon}. Meanwhile, the peaks at $Re>2500$ correspond to shear layer dynamics.

streamwise velocity components of the response and forcing modes corresponding to the maximum gain are also reported for each Reynolds number. At $Re=1000$, both the response and forcing modes present large structures developing at both the leading and trailing edges. The response mode extends to the wake region, while the forcing mode remains concentrated closer to the airfoil. Increasing the Reynolds number, the forcing and response modes lose support at the trailing edge and the modes concentrate in the shear layer region separating from the leading edge. Moreover, the forcing mode intensifies at the leading edge near the separation point and develops in the upstream region. This is due to the attenuation of viscous effects and the convective nature of these perturbations. The modal characteristic wavelength also becomes smaller with increasing Reynolds number. 

\begin{figure}
\centering
\includegraphics[width=0.9\textwidth]{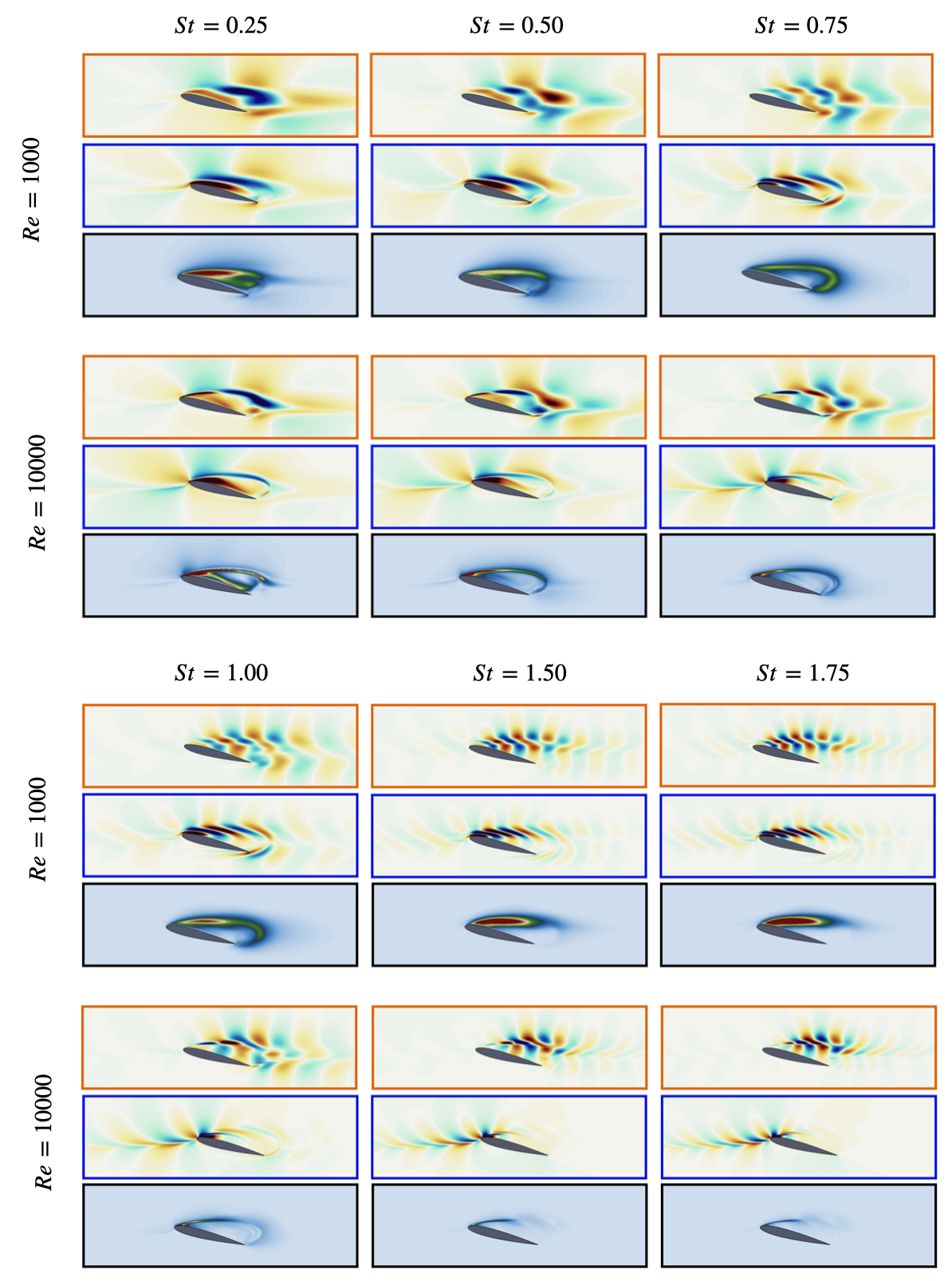}
\caption{Streamwise velocity component of the response modes (orange frame), streamwise velocity component of the forcing modes (blue frame) and magnitude of the wavemakers (black frame) shown for $\alpha=14^\circ$ at Reynolds numbers $Re=1000$ and $10000$ and $St\in[0.25;1.75]$. \label{fig:ModesAoA14}}
\end{figure}

The streamwise velocity components of the first response and forcing modes are shown together with the resolvent wavemaker in Figure~\ref{fig:ModesAoA14}, for $Re=1000$ and $10000$. The resolvent wavemaker is defined as the Hadamard, component-wise, product between forcing and response modes \citep{qadri2017frequency,skene2022sparsifying}:
\begin{equation}
    \hat{\mathbf{w}}=\hat{\mathbf{f}}\circ\hat{\mathbf{q}}.
\end{equation}
The resolvent wavemaker reveals regions that exhibit self-sustained mechanisms, thus regions where the response itself acts as a forcing. At the lowest frequency, we observe differences between the two cases in the response, forcing, and resolvent wavemaker structures, shown in orange, black, and blue frames, respectively. In particular, at $St=0.25$, we observe a thinning of the mode structures on the shear layer for the highest Reynolds number, which remains noticeable in the response mode structure up to $St\approx1$. Despite this difference in the shear layer, the response modes for both cases present similarities in the wake region for Strouhal numbers $0<St\leq1$. For higher $St$, the response mode structures at both the Reynolds number shift toward the shear layer regions presenting more similarities over this region.

For the considered range of Strouhal numbers, we in contrast observe a strong difference in the forcing modes. At the highest Reynolds number the forcing mode develops upstream, contrarily to the lower Reynolds number case for which the forcing mode structures are predominant in the shear layer region. This is also observed in the wavemaker field, which visualizes the overlap between the response and forcing modes. 

For $Re=10000$, we can observe again the thin elongated predominant structure over the shear layer, that weakens for higher $St$, contrarily to the $Re=1000$ case for which the wavemaker intensifies in the shear layer. For both $Re=1000$ and $10000$, we observe that the response and forcing modes for $St<1$ present structures on the wake, thus indicating wake dynamics, while for $St>1$ the modes shift toward the shear layer region, thus indicating shear layer dynamics.  

\begin{figure}
\centering
\includegraphics[width=\textwidth]{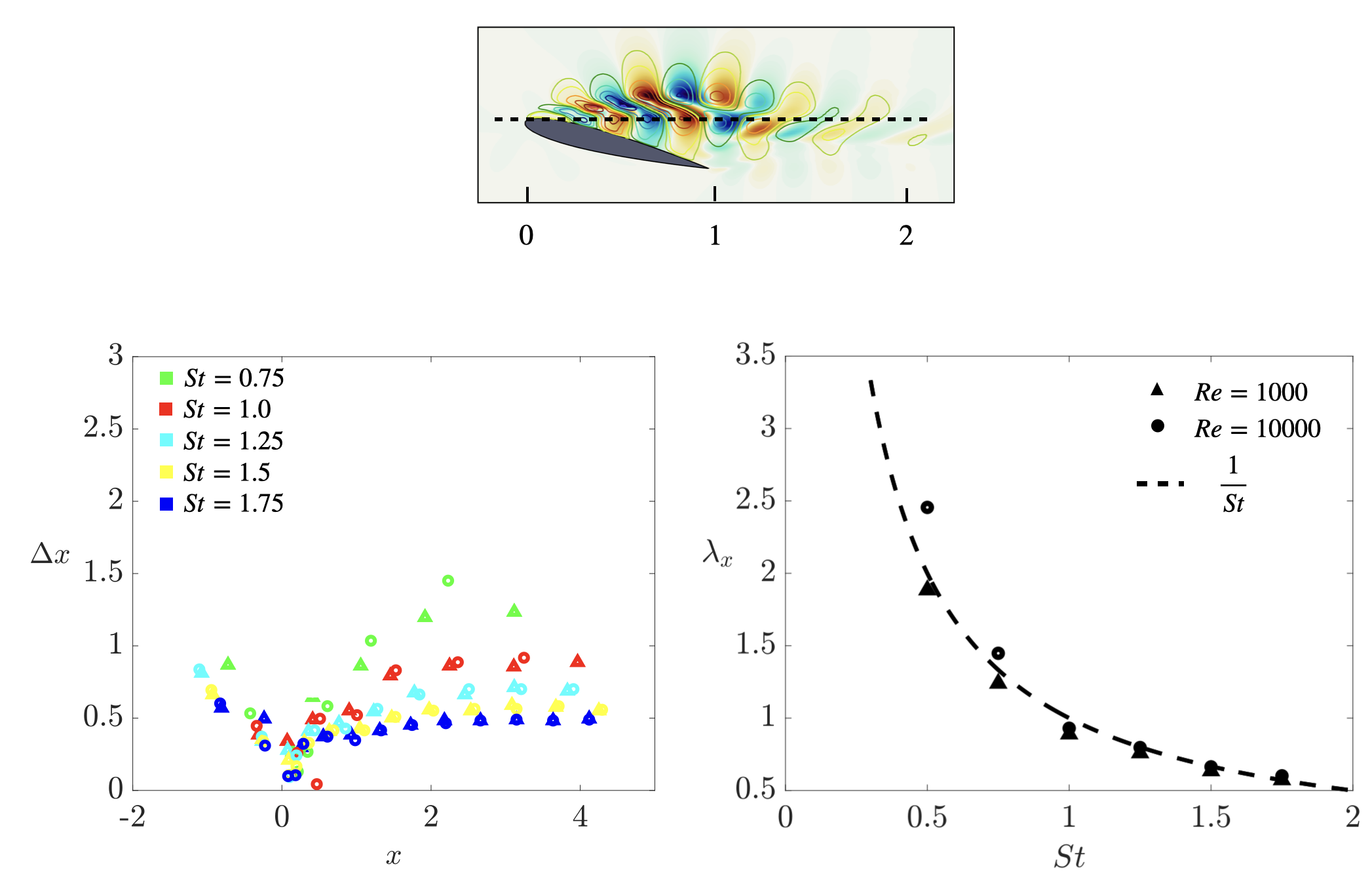}
\put(-460,195){\textit{(b)}}
\put(-220,195){\textit{(c)}}
\put(-305,300){\textit{(a)}}
\caption{(a) Streamwise distance between positions of the maximum peak of the response spatial structure over a line passing at the leading edge (black dashed line in (c)) and (b) characteristic wavelength as a function of the frequency. \label{fig:WaveLength}}
\end{figure}

Additionally, we observe that at the same frequency, the response modal structures match considerably well with each other for $St>0.5$. This is shown in figure  \ref{fig:WaveLength}.a, where the streamwise velocity response structure at $Re=1000$ is superposed with isocontours of the streamwise velocity response structure at $Re=10000$ for the same frequency. This similarity in the response structure comes from similarity in local wavelengths. To verify this, we consider the response variation over a line aligned along the streamwise direction and passing through the leading edge (black dashed line in Figure~\ref{fig:WaveLength}.a). The modal profile over this line is oscillatory, and the position of the maximum (or minimum) peaks can be tracked. The difference $\Delta x$ between the positions of the maximum peaks is shown in Figure~\ref{fig:WaveLength}.b as a function of the streamwise position, which captures the local streamwise wavelength.  We can observe that the local wavelengths of both Reynolds numbers follow the same trend and are in good agreement with each other. 

The maximum value of $\Delta x$ is reported as a characteristic wavelength in Figure~\ref{fig:WaveLength}.c as a function of the frequency. We notice that the values of $\lambda_x$ follow the $1/St$ curve, meaning that this characteristic wavelength is independent of the Reynolds number and only depends on the frequency. Moreover, this tells us that far from the airfoil, the modal structure corresponds to a wave of unitary phase velocity ($St\lambda_x=1$). Closer to the airfoil, instead, the local wavelength is lower, implying lower phase velocities. This is in agreement with previous studies even at higher Reynolds numbers \citep{boutilier2012separated}.

\section{Scaling of wake and shear layer dynamics}\label{sec:Normalizations}
Up to this point, we have focused on a specific angle of attack for analyzing the forcing modes, response modes, and gain distributions. Let us generalize how the shear and wake dynamics relate and shift from one another across the angles of attack and Reynolds numbers. To do so, we analyze the separated flow around a NACA0012 airfoil at an angle of attack of $\alpha=9^\circ$, and compare the results to those obtained for $\alpha=14^\circ$. We consider the case of spanwise wavenumber $\beta=0$.

\begin{figure} 
\centering
\includegraphics[width=\textwidth]{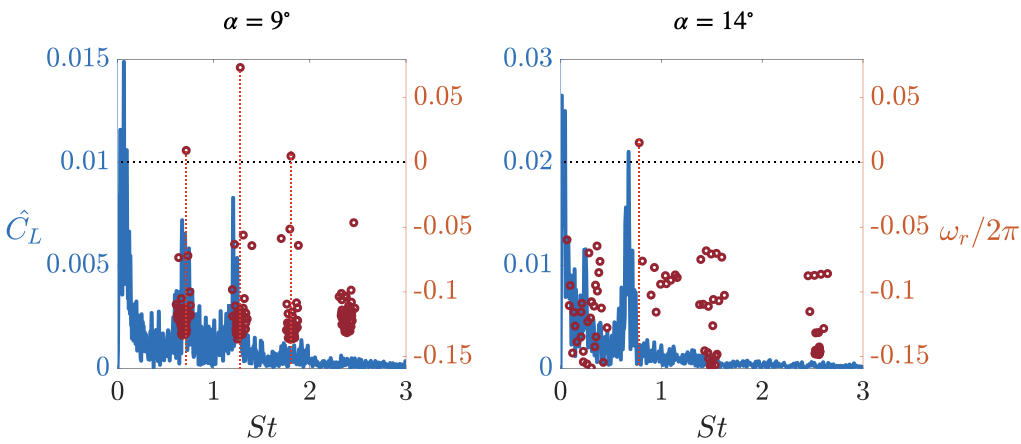}
\caption{ Lift spectra $\hat{C}_L$ shown together with the eigenvalues for $\alpha=9^\circ$ and $14^\circ$ at $Re=10000$.\label{fig:CL_OmegaAlpha}}
\end{figure}

The eigenvalues with the largest real component, together with the lift spectra at $Re=10000$ for the two angles of attack are shown in Figure~\ref{fig:CL_OmegaAlpha}. At $\alpha=9^\circ$, three eigenvalues have positive real component and reflect the peaks of the lift spectra. In particular, for both angles of attack, the eigenvalues with the largest real component relate to the largest amplitude peak in the lift spectra. The relative error between the shedding frequency peak $St_{C_L}$ and frequency of the least stable eigenvalue $St_{ev}$, computed by $|St_{C_L}-St_{ev}|/St_{ev}$ corresponds to $5.5\%$ for $\alpha=9^\circ$ and $13.4\%$ for $\alpha=14^\circ$. Therefore, by increasing the angle of attack, the difference between the frequency of the eigenvalues with the largest real component and the frequency peak of the lift coefficient also increases. This is due to stronger nonlinearities that are present in the unsteady flow at higher angles of attack. 

The largest real part of the positive eigenvalues at $\alpha=9^\circ$ is $\omega_r/2\pi=0.073$, which suggests that we consider dynamics over a timescale shorter than $t_\gamma=13.7$. This allows us to use the discount parameter corresponding to $t_\gamma=5$, as in the previous section.

\begin{figure} 
\centering
\includegraphics[width=0.9\textwidth]{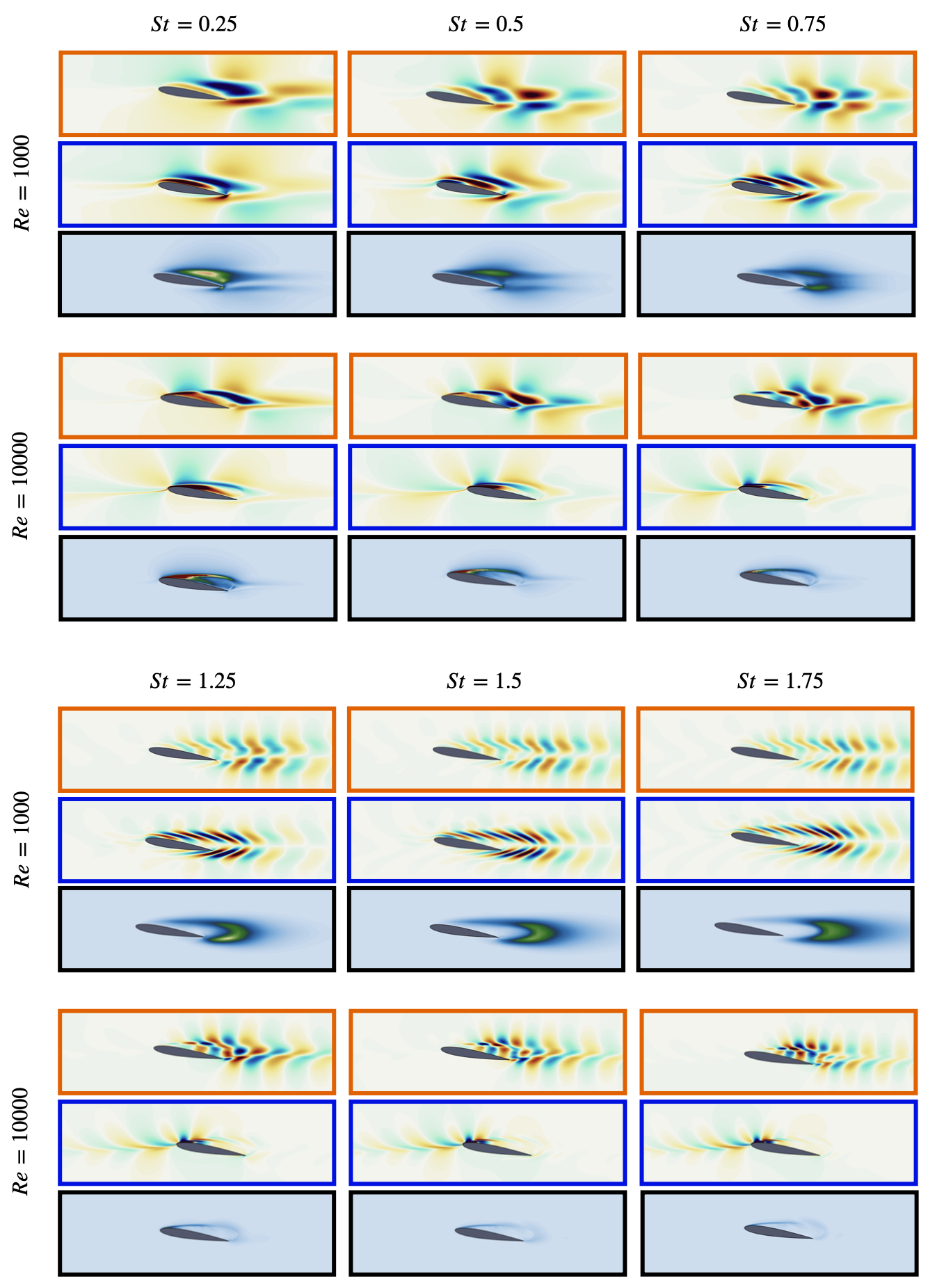}
\caption{Streamwise velocity component of the response modes (orange frame), streamwise velocity component of the forcing modes (blue frame) and magnitude of the wavemakers (black frame) shown for $\alpha=9^\circ$ at $Re=1000,10000$ and $St\in[0.25;1.75]$. \label{fig:AoA_9-Modes}}
\end{figure}

The streamwise velocity component of the response and forcing modes, along with the wavemaker of the first singular mode for $\alpha=9^\circ$ at $Re=1000$ and $10000$, are shown in Figure~\ref{fig:AoA_9-Modes}. We compare the response and forcing modes at the two angles of attack (see also Figure~\ref{fig:ModesAoA14} for the $\alpha=14^\circ$ case). Although not reported here, we observe that the characteristic streamwise wavelength $\lambda_x$, introduced in Sect.~\ref{sec:ResultResolvent1} for the $14^\circ$ case, exhibits the same characteristics for $\alpha=9^\circ$ and follows $\lambda_x=1/St$. Moreover, similarities are observed in the forcing and response modes between the two angles of attack. 

At $Re=1000$, both the response and forcing modes are similar between $\alpha=9^\circ$ and $14^\circ$ at low frequencies, up to $St\approx0.75$. 
 The structure of both cases is mainly concentrated in the downstream part of the recirculation region, with a dominating wake structure. However, some differences can be seen in the wavemaker. For the $\alpha=9^\circ$, indeed, the wavemaker vanishes close to the suction side, while this is not the case for the $14^\circ$ case, for which a high magnitude of the wavemaker is present at the separation point. For higher frequencies, $St\geq0.75$, the response and forcing modes of the two angles of attack are different. For $\alpha=9^\circ$, the structures persist downstream in the wake region but with reduced intensity. In contrast, for $\alpha=14^\circ$, the modal structures gradually shift toward the shear layer. This behavior is likely due to the fact that, at $Re=1000$, the separated flow around the airfoil at $\alpha=9^\circ$ is still two-dimensional, while at $\alpha=14^\circ$ the flow is three-dimensional. 

The comparison between the resolvent modes across $\alpha=9^\circ$ and $14^\circ$ at the higher Reynolds number, $Re=10000$, also reveals similarities between the two angles of attack, but again only at the lower frequencies, up to $St\approx1$. 
For the $9^\circ$ case, the shift of the response from the wake to the shear layer region is not as clear as for the $14^\circ$ case. 
At $\alpha=9^\circ$ and $St\geq 1$, the modes present structures in both the wake and the shear layer. In particular, wake structures persist for higher frequencies compared to the $14^\circ$ case.
This is because wake dynamics are governed by the angle of attack. In fact, the wake characteristic frequency depends on the width of the wake \citep{roshko1954development}, which becomes thinner as the angle of attack decreases and can thus support higher frequencies compared to higher angles of attack.

\begin{figure} 
\centering
\includegraphics[width=\textwidth]{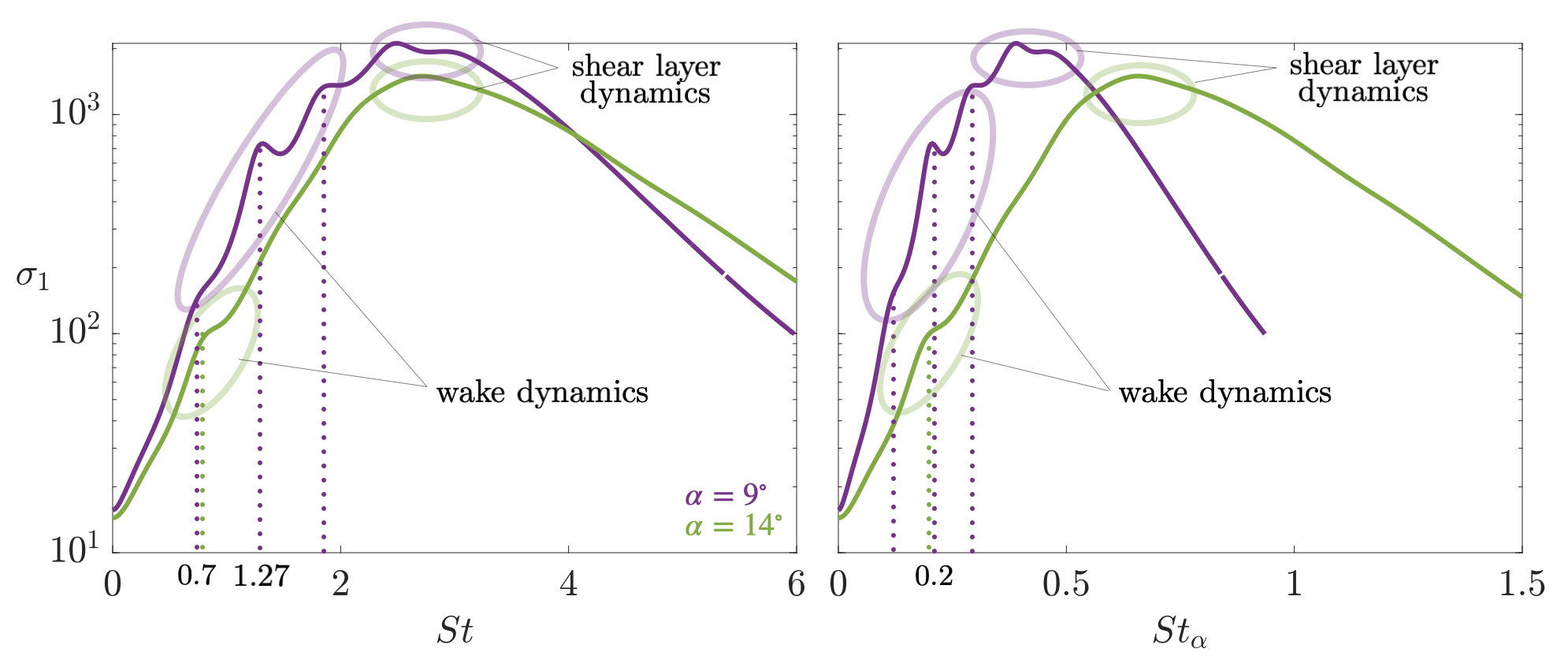}
\put(-462,190){$(a)$}
\put(-222,190){$(b)$}
\caption{Gain distributions of the first mode for $\alpha=9^\circ$ and $14^\circ$ at $Re=10000$ over (a) Strouhal number $St$ based on the chord and (b) Strouhal number $St_\alpha$ based on the front facing area. Dotted lines indicate the frequency associated with the positive eigenvalues (Figure~\ref{fig:CL_OmegaAlpha}).\label{fig:Alpha_9_14_Sigma}}
\end{figure}

The gain variation of the first mode with respect to $St$ and $St_\alpha$, where 
  \begin{equation}
     St_\alpha=\frac{\omega c \sin{\alpha}}{2\pi U_\infty}=St\sin{\alpha},
 \end{equation}
 are shown in Figure~\ref{fig:Alpha_9_14_Sigma}.
The figure annotates the regions of the gain distribution that correspond to wake or shear layer dynamics. The energy amplification at $\alpha=9^\circ$  is higher compared to $\alpha=14^\circ$, and exhibits pronounced local maximum. This is due to the eigenvalues presented in Figure~\ref{fig:CL_OmegaAlpha}. The higher $\omega_r$ and multiple eigenvalues for $\alpha=9^\circ$ translate in a higher gain and multiple ``bumps" compared to the $\alpha=14^\circ$ case, due to the proximity of the integration path to the eigenvalues. In Figure~\ref{fig:Alpha_9_14_Sigma}.a, we observe that the most pronounced relative peak for $\alpha=9^\circ$, at $St\approx 1.3$, corresponds to the largest eigenvalues and highlights wake dynamics. This occurs at a higher frequency compared to the $\alpha=14^\circ$ local peak, around $St\approx 0.7$. On the contrary, for the two angles of attack, the maximum at higher frequencies, correspondent to the shear layer dynamics, occurs at about the same frequency, $St\approx 2.7$.

When looking at the gain variation with respect to $St_\alpha$, Figure~\ref{fig:Alpha_9_14_Sigma}.b, we observe that the wake dynamics is predominant at the same frequency for the two angles of attack. In particular, the most pronounced low-frequency peaks occur at $St_\alpha\approx 0.2$ \citep{fage1927flow}, indicating phenomena induced by the frontal wing height. However, we can see that the most energetic shear layer dynamics at the two angles of attack does not occur at the same frequencies when considering frequencies based on the angle of attack. Moreover, in the previous section, we observed that the frequency of maximum gain increases with the Reynolds number (see Figure~\ref{fig:SigmaModesAoA14}). These two observations highlight dynamics within the shear layer which depend on the Reynolds number but not on the angle of attack. This can be also observed in the results presented by \cite{yeh2019resolvent}, where the flows at Reynolds number $Re=23000$ and angles of attack $\alpha=6^\circ$ and $9^\circ$ are investigated. The shear dynamics was observed not to be influenced by the angle of attack the angle of attack, with maximum peak at $St\approx 4.8$ for both $\alpha$ for their cases.
\begin{figure} 
\centering
\includegraphics[width=\textwidth]{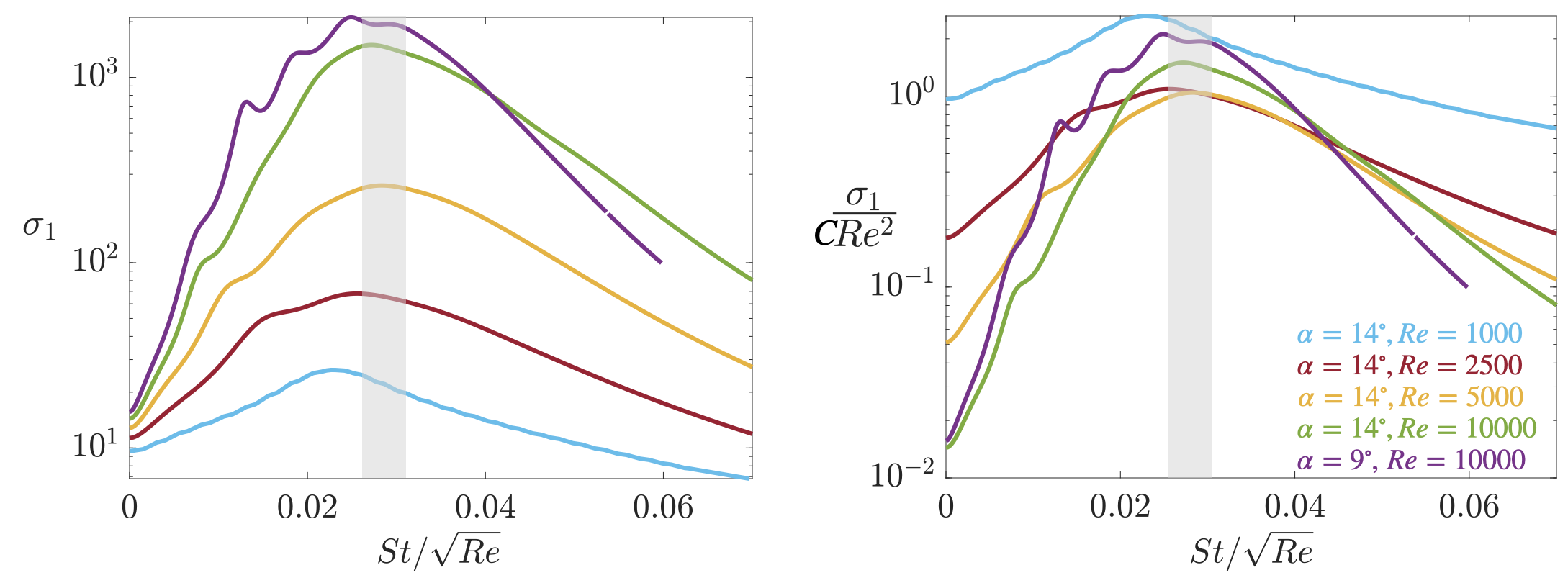}
\put(-462,180){$(a)$}
\put(-222,180){$(b)$}
\caption{(a) Frequency and (b) gain normalization for various angles of attack and Reynolds number.\label{fig:Normalization}}
\end{figure}

When the boundary layer separates without reattaching to the airfoil, it exhibits a characteristic frequency that scales like that of a free shear layer \citep{ho1984perturbed,kotapati2010nonlinear}. The frequency depends on the boundary layer thickness at the separation point \citep{bloor1964transition}. 
We show the gain variation over the normalized frequency based on the laminar boundary layer thickness $\delta \approx 1/\sqrt{Re}$ \citep{bloor1964transition}
\begin{equation}
  St_{Re}=\frac{\omega c \delta}{2 \pi U_\infty}=\frac{St}{\sqrt{Re}}
\end{equation}
in Figure~\ref{fig:Normalization}.a. We observe that the maximum gains occur at a similar normalized frequency. In fact, we find that the frequency of maximum gain and the corresponding streamwise characteristic length of the shear layer dynamics occur at 
\begin{equation}\label{eq:Normal}
    St\approx0.027\sqrt{Re}, \quad \lambda_x\approx 37/\sqrt{Re}.
\end{equation}
The value of the normalized Strouhal number is in agreement with the range of $St/\sqrt{Re}\in [0.02;0.03]$ proposed by \cite{zaman1991control} as the range of effective excitation frequencies in acoustic control of flow over airfoils \citep{yarusevych2003effect,gencc2016acoustic}. \cite{ho1984perturbed} also report a similar value of the normalized shear layer frequency, close to $0.03$, for a laminar flow, when normalized by the momentum thickness and the average velocity across the shear layer \citep{ho1984perturbed, kotapati2010nonlinear}. Results from \cite{yeh2019resolvent} are also in agreement with the most amplified frequency related to shear layer dynamics ($St/\sqrt{Re}\approx4.8/\sqrt{23000}\approx0.032$). Experimental results from \cite{klewicki2024footprint} at higher Reynolds numbers also fall in the same range ($St/\sqrt{Re}\approx 0.021$ for $Re=2\times 10^4$ and $St/\sqrt{Re}\approx 0.027$ for $Re=4-8\times 10^4$). Their study examines the flow around a wing of aspect-ratio $3$ with mounted walls, and the lower value of $St/\sqrt{Re}$ at $Re=2\times 10^4$ is likely due to the effects of the wall that at the lower Reynolds number cause a higher three-dimensionality of the flow and a stabilization of the shear layer at the extremes.

 It is worth noticing that the $Re=1000$ case does not align with the normalization. This is due to the fact that, at this Reynolds number, the dynamics (and the peak) are associated with wake dynamics rather than shear layer ones, as discussed in Sect.~\ref{sec:FiniteHorizon}.

The energy amplification is scaled by $Re^2$ in Figure~\ref{fig:Normalization}.b, showing an almost quadratic variation of the amplification energy with respect to the Reynolds number. The distributions of $\sigma_1$ are also scaled by constant $C$ that scales the peaks to be close to $1$. In this case, we have used $C=10^{-5}$. Previous works show the maximum gain to quadratically depend on the Reynolds number in planar flows such as plane Poiseulle, Couette flow, and Blasius boundary layer \citep{schmid2002stability}, but also in accelerating-decelerating flows \citep{linot2024laminar}, and oscillatory flows \citep{xu2021non}.

\section{Conclusions}
We provide a comprehensive analysis of the behavior of separated flows over an airfoil under spanwise homogeneous conditions. The study explores Reynolds numbers spanning one to two orders of magnitude higher than earlier work, highlighting two distinct dynamics at play and documenting their characteristic frequencies.

To do so, we emploied biglobal resolvent analysis and investigate the effects of the Reynolds number $Re=1000$, $2500$, $5000$ and $10000$ on separated flow around a NACA0012 airfoil at $14^\circ$ angle of attack. To compute the base flows, we performed direct numerical simulations for $Re=1000$ and $2500$, and wall-resolved large eddy simulations for $Re=5000$ and $10000$. At these Reynolds numbers, the flow is three-dimensional, presenting spanwise periodic structures at $Re=1000$ and $2500$ whose wavelengths decrease as the Reynolds increases and the flow becomes more chaotic. The two-dimensional base flows were obtained by performing a time- and spanwise-average of the unsteady flow.

We observed that the recirculation region shortens from $Re=1000$ to $Re=2500$, before elongating again as the Reynolds number increases further.  Additionally, for $Re\geq 2500$, a secondary recirculation region emerges on the suction side of the airfoil and remains present at higher Reynolds numbers. The energy spectra evaluated at four streamwise locations along the shear layer, showed high-frequency contents at specific cross-stream locations, and at higher frequencies when increasing the Reynolds number. 

Our results were organized in two parts. In the first part, the results of the resolvent analysis were examined with respect to the different parameters: discount parameter, spanwise wavenumber, and frequency.  Varying the discount parameter allowed us to consider the dynamics over different timescales. The results showed that, at short timescales, shear layer dynamics are the most energetic, while at longer timescales wake dynamics prevail. Three-dimensionality, investigated by varying the spanwise wavenumber, also seems to be effective at long timescales and to be sustained by wake dynamics, at low frequencies. In contrast, shear layer dynamics, that occur at high frequencies, mainly remain two-dimensional. This reflects the vortex roll-up of the shear layer, which is a two-dimensional phenomenon.     

The effects of the Reynolds number at a fixed timescale were then investigated, comparing modal structures and the energy gain variations. We analyzed the predominance of shear layer and wake dynamics with respect to the frequency, identifying the transition point at $St\approx 1$. We observed similarities in the forcing and response mode structures across different Reynolds numbers. The characteristic streamwise wavelength was found to be independent of the Reynolds number, depending only on the frequency and ensuring a unitary characteristic phase velocity, despite differences in energy amplification.

In the second part, we compared the results with a different angle of attack, still focusing on the shear layer and wake dynamics. While wake dynamics are influenced by the angle of attack, shear layer dynamics depend solely on the Reynolds number. The main frequencies that characterize the two different dynamics approach each other when decreasing the angle of attack at a constant Reynolds number, while they separate when increasing the Reynolds number at a constant angle of attack. Normalizing the Strouhal number by the Reynolds number ($St/ \sqrt{Re}$) highlights the shear layer scaling, with maximum energy amplification occurring at $St\approx0.027\sqrt{Re}$, consistent with prior studies. Moreover, the energy amplification scales quadratically with $Re$. 

This study reveals the dominant wake and shear layer dynamics, emphasizing their dependence on the Reynolds number and angle of attack. The identified scalings and trends bridge gaps in understanding transitional flow regimes. These insights are useful for improving predictions and control strategies for flows at even higher Reynolds numbers.

\section*{Funding}
This work was supported by the U.S. Army Research Office (Grant Number W911NF-21-1-0060) and the U.S. Air Force Office of Scientific Research (Grant Number FA9550-21-1-0174)

\section*{Acknowledgments}
This work used computational and storage services associated with the Hoffman2 Shared Cluster provided by UCLA Institute for Digital Research and Education’s Research Technology Group. LVR and KT thank Charles Klewicki for insightful discussions.

\appendix
\section{Effects of the discount parameter} \label{Appendix1}
\begin{figure} 
\centering
\includegraphics[width=0.5\textwidth]{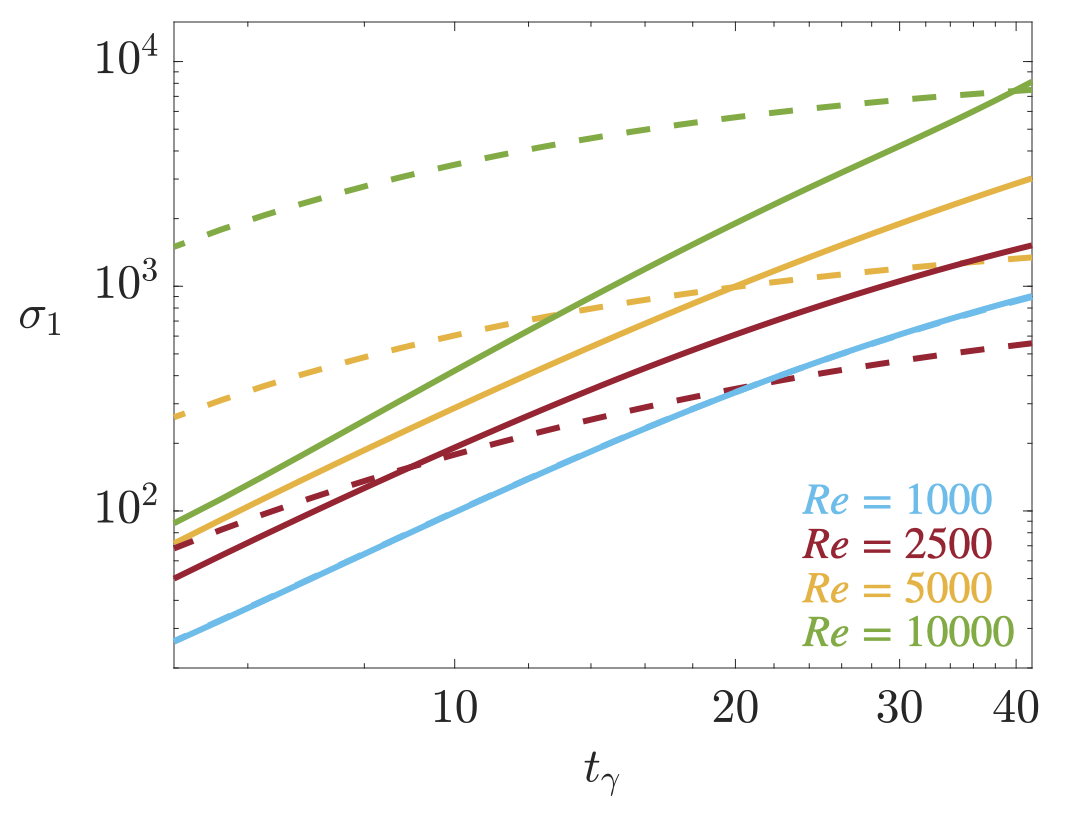}
\caption{Variation of the first singular value $\sigma_1$ over time $t_\gamma$ at the frequencies of the maximum gain over short and long timescales. (\full) indicates the wake mode frequency (lower frequency peak) and (\dashed) indicates the shear layer mode frequency (higher frequency peak)    \label{fig:Sigma1Discount}}
\end{figure}

In this Appendix, we report further analysis on the effect of the discount parameter $\gamma$. In Sect. \ref{sec:FiniteHorizon}, it has been shown that the dominance of the wake and shear layer modes changes with $t_\gamma=2\pi/\gamma$. In particular, at short timescale, higher energy amplifications occur at high frequency, in the shear layer, while at long timescale, wake dynamics arising at low frequency show higher energy amplification. Figure~\ref{fig:Sigma1Discount} shows the energy gain of the wake and shear layer modes, at $St_W$ and $St_S$, respectively, over $t_\gamma$. The energy gain of the two modes increases over time with different slopes. In particular, we can observe that the wake mode energy increase is steeper compared to the shear layer mode, while the shear layer mode seems to tend toward an asymptotic plateau. 
From this plot, we can see that the time at which the wake mode prevails over the shear layer mode increases with the Reynolds number.

\begin{figure} 
\centering
\includegraphics[width=\textwidth]{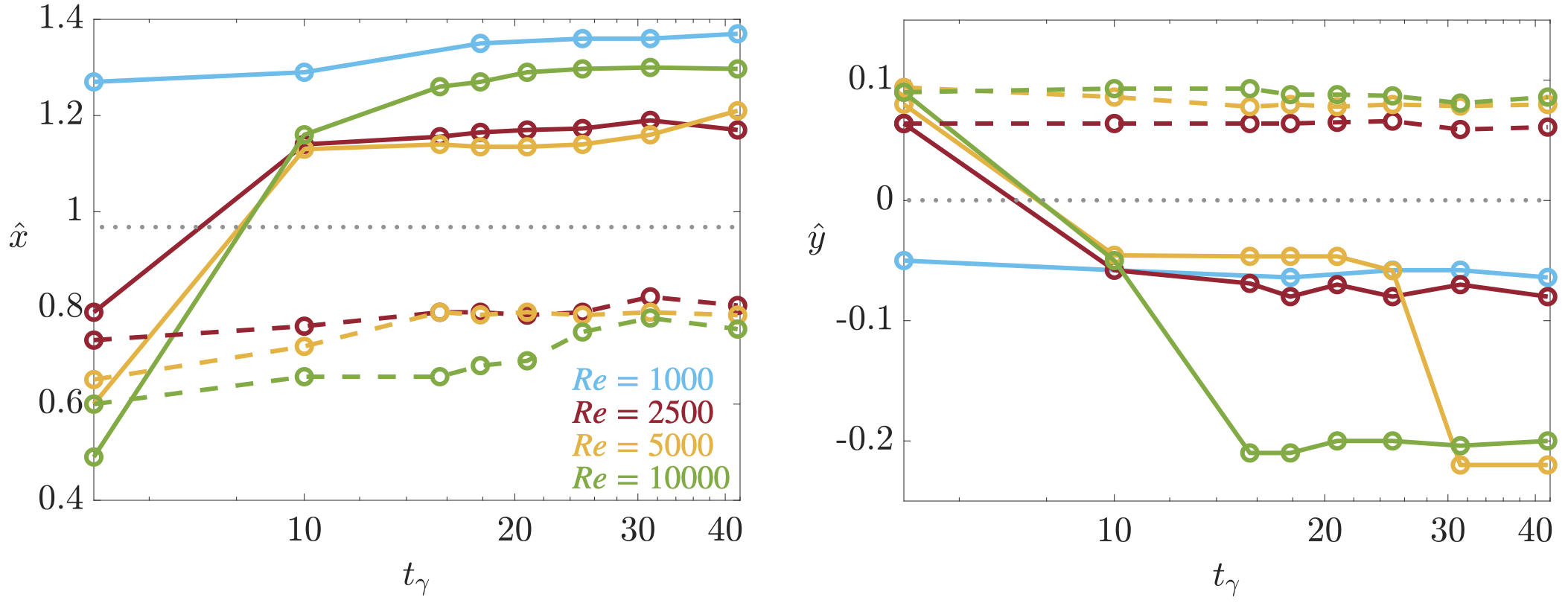}
\put(-294,117){\small{wake region}}
\put(-314,107){\small{shear layer region}}
\put(-52,115){\small{wake region}}
\put(-72,125){\small{shear layer region}}
\caption{Streamwise and cross-stream position of the maximum kinetic energy of the shear response mode (\dashed) and wake response mode (\full) over time $t_\gamma$.  \label{fig:DiscountMaxRespPosition}}
\end{figure}
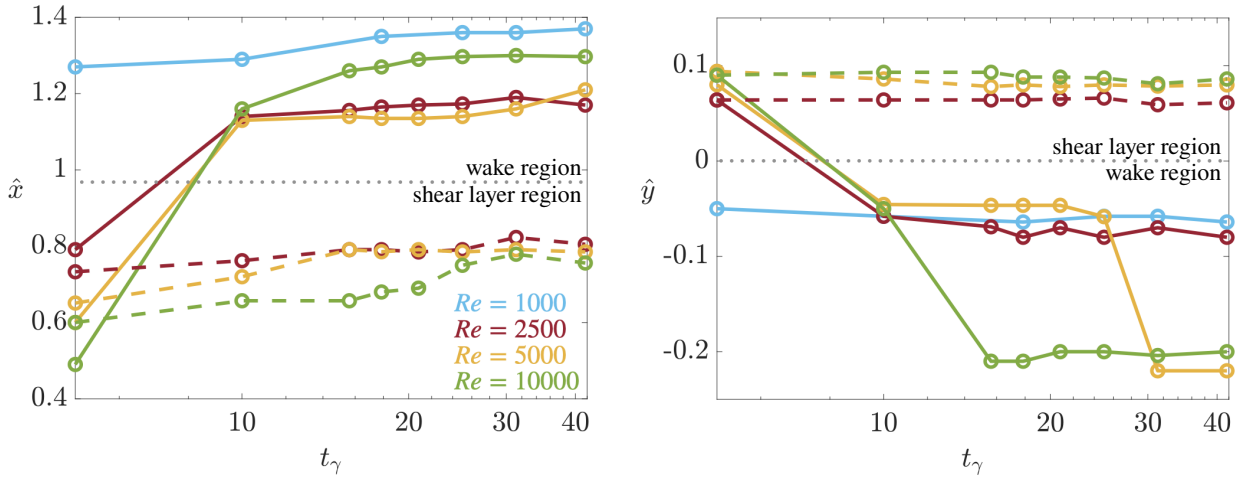

The variation in time of the response modes can be investigated also by tracking the streamwise and cross-stream position of the maximum kinetic energy of the response mode. This is shown in Figure~\ref{fig:DiscountMaxRespPosition}, considering
\begin{equation}
    \{\hat{x},\hat{y}\}=\text{arg}\max_{x,y}||\hat{\mathbf{q}}_{\mathbf{u}}||_2(x,y),
\end{equation}
where $\hat{\mathbf{q}}_{\mathbf{u}}=(\hat{q}_{u_x},\hat{q}_{u_y},\hat{q}_{u_z})$. 
From these plots, we observe that for the cases in which the shear layer supports energetic modal structures ($Re\geq2500$), the location of the shear layer mode's maximum intensity remains almost unchanged. Interestingly, in these cases, the wake mode is most intense in the shear layer region over short times before eventually shifting toward the wake region. The streamwise and cross-stream location of the shear mode and wake mode over short timescale show good agreement with the region of high amplitude in the energy spectra contour shown in Figure~\ref{fig:ProbeBOX}. At $Re=1000$, the location of the wake mode's maximum intensity remains almost constant over time. Additionally, the streamwise position of the wake mode at long timescales is correlated to the length of the base flow recirculation region (see Figure~\ref{fig:baseflow}.e). The response mode at $Re=1000$ is most intense farther downstream compared to the higher Reynolds number cases, as its recirculation region is the most elongated. Moreover, the response modes at $Re=2500$ and $5000$ are most intense at a similar streamwise location as their recirculation regions have comparable extensions.

\section{Spanwise wavenumber effects on the response modal structures}
\begin{figure} 
\centering
\includegraphics[width=\textwidth]{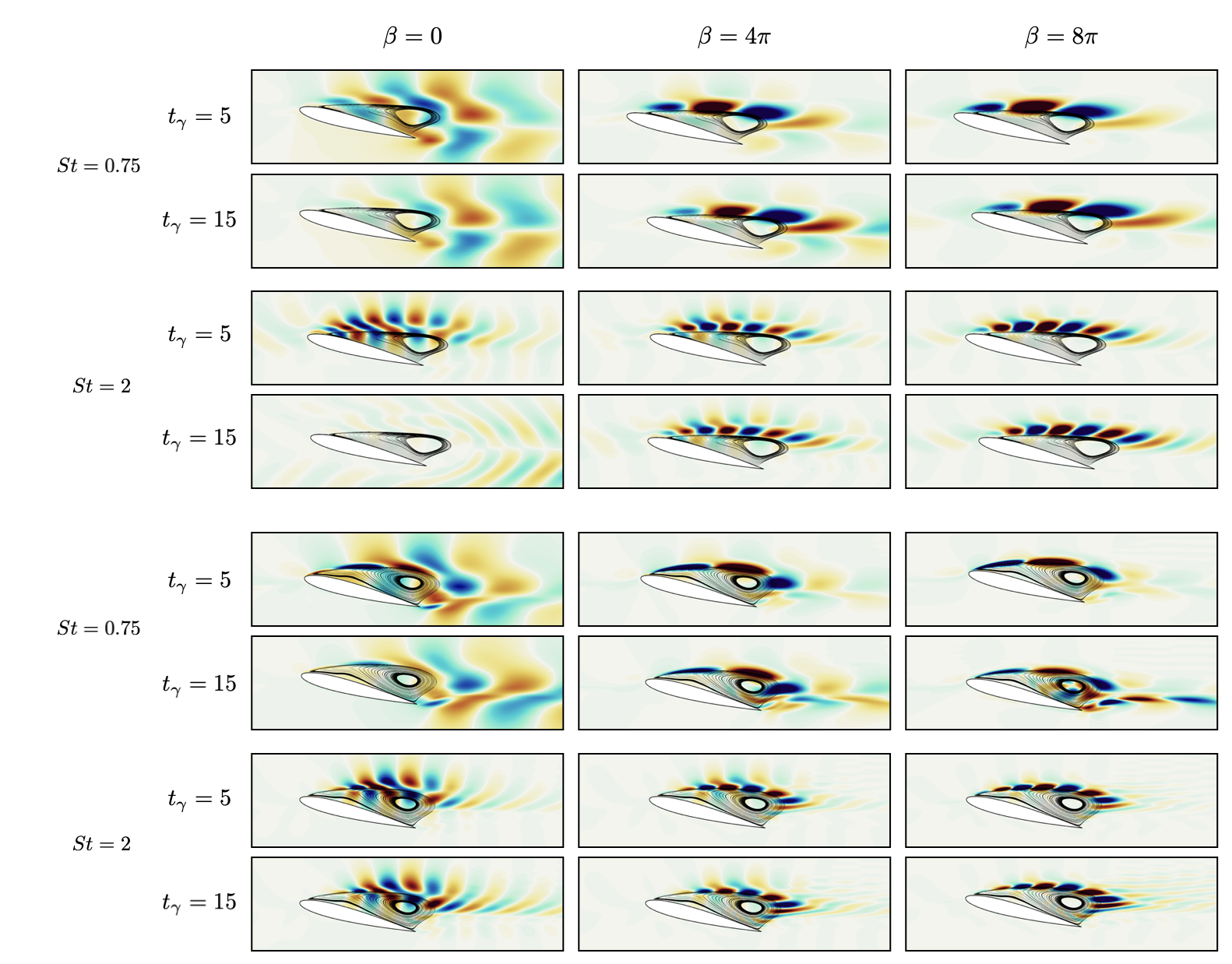}
\put(-468,260){{$Re=1000$}}
\put(-468,83){$Re=10000$}
\caption{Streamwise velocity response contours at $St=0.75, 2$ and $\beta=0,4\pi$ and $8\pi$ for $Re=1000,\;10000$ and $t_\gamma=5$ and $15$,  superposed to base flow velocity streamlines within the recirculation region. \label{fig:BetaModes} }
\end{figure}

In Sect. \ref{sec:BetaEffects} we have presented the effects of the spanwise wavenumber $\beta$ on the gain variation over the frequency. In this Appendix, we report the changes in the response modal structures with respect to $\beta$. The effect of $\beta$ on the response mode structures is shown in Figure~\ref{fig:BetaModes} for $Re=1000$ and $10000$, at frequencies $St=0.75$ and $2$ and two different timescales. The streamlines of the base flow are also plotted to highlight the recirculation region.

We first consider the lower frequency, $St=0.75$. Over $t_\gamma=5$ and at $\beta=0$, the response mode structure develops in the wake region for both Reynolds numbers. The structure presents alternating oblique structures characteristic of the streamwise velocity component of oscillating modes.  As $\beta$ increases, these elongated structures gradually evolve into alternating concentrated structures located in the shear layer. This transition occurs because smaller structures require stronger mean shear for amplification \citep{yeh2019resolvent, skene2022sparsifying}. Over $t_\gamma=15$ and $Re=1000$, the behavior observed over the shorter timescale persists. However, over $t_\gamma = 15$ and $Re = 10000$, the mode becomes concentrated in the recirculation region and the trailing-edge shear layer region at the highest $\beta$.

Now, let us consider the higher frequency, $St=2$.  Over $t_\gamma=5$, the response mode for both Reynolds numbers gradually transitions from alternating oblique structures to alternating concentrated structures within the shear layer region. Over  $t_\gamma=15$ and at $Re=1000$, a mode switching occurs and structures emerge in the shear layer for higher $\beta$ values, which are absent at $\beta = 0$.  Over  $t_\gamma=15$ and at $Re=10000$ the mode structure does not significantly change compared to the short timescale.

 \bibliographystyle{abbrv}  
\bibliography{main} 
\end{document}